\documentclass[lettersize,journal]{IEEEtran}
\usepackage{amsmath,amsfonts}
\usepackage{algorithmic}
\usepackage{algorithm}
\pdfoutput=1
\usepackage{array}
\usepackage[caption=false,font=normalsize,labelfont=sf,textfont=sf]{subfig}
\usepackage{textcomp}
\usepackage{stfloats}
\usepackage{url}
\usepackage{verbatim}
\usepackage{graphicx}
\usepackage{cite}
\usepackage{svg}
\usepackage{makecell}
\usepackage{booktabs} 
\usepackage{makecell} 
\usepackage{multirow} 
\usepackage{mathptmx} 
\usepackage{hyperref} 

\usepackage{bm}

\usepackage{newtxtext,newtxmath} 

\begin{document}

\title{3DMeshNet: A Three-Dimensional Differential Neural Network for Structured Mesh Generation}

\author{
    Jiaming Peng, 
    Xinhai Chen,\IEEEauthorrefmark{1} 
    Jie Liu
    \IEEEauthorblockA{
        \\
        Laboratory of Digitizing Software for Frontier Equipment, \\
        Science and Technology on Parallel and Distributed Processing Laboratory,\\
        National University of Defense Technology, Changsha 410073
    }\\
    \thanks{\IEEEauthorrefmark{1}Xinhai Chen, Corresponding Author, Email: chenxinhai16@nudt.edu.cn}
}


\renewcommand\thead[2][c]{%
  \def\arraystretch{1}%
  \makecell[#1]{#2}%
}

\maketitle

\begin{abstract}
Mesh generation is a crucial step in numerical simulations, significantly impacting simulation accuracy and efficiency. However, generating meshes remains time-consuming and requires expensive computational resources. In this paper, we propose a novel method, 3DMeshNet, for three-dimensional structured mesh generation. The method embeds the meshing-related differential equations into the loss function of neural networks, formulating the meshing task as an unsupervised optimization problem. It takes geometric points as input to learn the potential mapping between parametric and computational domains. After suitable offline training, 3DMeshNet can efficiently output a three-dimensional structured mesh with a user-defined number of quadrilateral/hexahedral cells through the feed-forward neural prediction. To enhance training stability and accelerate convergence, we integrate loss function reweighting through weight adjustments and gradient projection alongside applying finite difference methods to streamline derivative computations in the loss. Experiments on different cases show that 3DMeshNet is robust and fast. It outperforms neural network-based methods and yields superior meshes compared to traditional mesh partitioning methods. 3DMeshNet significantly reduces training times by up to 85\% compared to other neural network-based approaches and lowers meshing overhead by 4 to 8 times relative to traditional meshing methods.
\end{abstract}


\section{Introduction}
\IEEEPARstart{N}umerical simulation has become increasingly important in various fields of natural science research, promoting the development of modern applied scientific research and engineering technology \cite{chawner2019progress,hayase2015numerical}. These simulations typically require the use of meshes as a fundamental geometric framework to facilitate the analysis of physical phenomena via finite element methods or other numerical techniques. Mesh generation, defined as dividing a continuous computational domain into meshes or elements for further numerical solutions, is a critical step in engineering simulations\cite{chen2021mve}. The quality of the generated mesh has a significant impact on the accuracy of numerical simulations\cite{slotnick2014cfd}. Given the significant impact of mesh quality on these aspects, the rapid generation of high-quality meshes has become a primary concern.

Structured meshes are widely used in numerous simulations requiring precise element alignment for high-efficiency and high-precision\cite{owen2015evaluation}. They present multiple benefits: (1) Structured meshes do not require additional storage to connect mesh elements, resulting in higher efficiency in data access. (2) Their structured topological nature facilitates easy control over mesh cell sizes. (3) The regularity of their structure facilitates the implementation of convolutional and pooling operations in neural networks \cite{logg2009efficient}. Motivated by these benefits, extensive research has been done on structured mesh generation\cite{chen2022improved,chen2022mgnet}. Algebraic methods calculate mesh interior points through algebraic interpolation, employing techniques like polynomial and Transfinite Interpolation (TFI)\cite{allen2008towards}. This approach is straightforward and enables rapid mesh generation. However, it can lead to mesh degradation in areas with complex geometries, resulting in issues such as intersections. To mitigate these problems, Partial Differential Equation (PDE) methods are commonly used to generate meshes for arbitrary boundaries\cite{chen2022mgnet}. Given the boundary values, PDE methods pose mesh generation as a boundary value problem governed by differential equations. Meshes created using PDE methods are of higher quality but require many computational resources and are time-consuming\cite{turner2017high}. Thus, there is a significant demand for further research and development of a rapid and robust structured mesh generation method that balances speed and mesh quality.

In recent years, deep neural networks (DNNs) have achieved remarkable progress in fields such as flow-field prediction\cite{obiols2020cfdnet}, mesh quality determination\cite{chen2021mve}, and other physical problems\cite{lu2019deeponet}. A few methods have been developed to incorporate neural networks into mesh generation tasks. Jilani et al. \cite{jilani2009adaptive} proposed a two-dimensional(2D) initial mesh generated based on a Self-organizing map neural network for finite element analysis and calculation of mesh self-adaptive parameters, followed by adaptive tuning of the initial mesh using the Self-organizing map neural network. Liu et al.\cite{liu2023ispliter} introduced the ISpliter method, combining neural networks with segmentation lines to transform B-rep geometrical shapes into unstructured meshes and facilitating neural network training through recursive data generation. Soman S\cite{soman2023faster} employed Conditional Generative Adversarial Networks to ascertain the coordinates of mesh nodes for generating 2D and three-dimensional (3D) unstructured meshes. However, these methods often rely on large training datasets, which can be costly and complex to obtain, particularly for geometries of intricate shapes. Methods like MGNet\cite{chen2022mgnet}, which utilize unsupervised learning for mesh generation, address the challenge of extensive manual training data requirements but still encounter problems such as unstable convergence and long training times. Additionally, pioneer researchers primarily focus on generating structured meshes in 2D, and developing neural network-based methods for 3D meshes remains an open problem.

In this paper, we introduce 3DMeshNet, a novel intelligent method for 3D structured mesh generation. 3DMeshNet employs a neural network embedded with 3D elliptic PDEs to learn the meshing rules for 3D mesh partitioning. Specifically, a well-designed neural network is developed to approximate the potential mapping between parametric and computational domains. The 3D elliptic PDEs and surface fitting items are embedded in the loss function, acting as penalty terms to guide the optimization of network parameters. After suitable training, 3DMeshNet can efficiently generate meshes of any specified size for the corresponding geometry using forward propagation. To reduce training time, improve surface fitting accuracy, and enhance model convergence stability, we employed finite differences (FD), loss function reweighting, and gradient projection to improve the model from three aspects. Finite differences reduce computational costs and accelerate training by efficiently computing derivatives. Loss function reweighting improved surface representation by adjusting loss terms using multi-task learning. Gradient projection reconciled gradients from different loss items. Our experiments on 2D and 3D geometries demonstrate that 3DMeshNet can effectively balance meshing overhead and mesh quality, producing high-quality meshes for complex shapes where PINN methods might struggle to maintain mesh quality. 3DMeshNet produces mesh quality that surpasses existing approaches, achieving a meshing overhead reduction of 4 to 8 times compared to TFI methods and decreasing training times by up to 85\% relative to other neural network-based methods. To the best of our knowledge, this is the first work to introduce PINNs for 3D mesh generation.

The remainder of this paper is organized as follows: Section \ref{Related work} reviews related work on mesh generation methods. Section \ref{method} provides a detailed description of the proposed 3DMeshNet method. Section \ref{Experimental} presents experimental results comparing different methods on two 2D geometries and four 3D geometries. Lastly, Section \ref{Conclusion} concludes the paper and discusses future work.

\section{Related work}
\label{Related work}

Structured meshes provide simplicity, making them widely used in numerous simulations that necessitate precise alignment of elements as dictated by analytical requirements\cite{lintermann2021computational}. The demand for precision has also stimulated the development of mesh generation methods\cite{chen2022mgnet,chen2022improved}. The methods for generating structured meshes include conformal mapping, algebraic methods, and PDE methods\cite{sarrate2014unstructured}. Algebraic methods typically employ various interpolation techniques or special functions for mesh creation\cite{bern2000mesh}. Conformal mapping, which employs angle-preserving transformations derived from complex function theory, facilitates the generation of 2D meshes by altering the computational domain. However, expanding these methods to 3D mesh generation presents significant challenges. PDE methods, on the other hand, generate meshes by solving elliptic, parabolic, or hyperbolic equations within a specified transformation domain\cite{babuska1995modeling}. Although the algebraic method offers simplicity and rapid mesh creation, it may lead to mesh degradation in areas of complex geometry or cause mesh elements to overlap or breach boundaries. Structured meshes generated using the differential equation method are of high quality but require high computational effort.

Artificial intelligence represented by neural networks has been vigorously developed in recent years\cite{gawlikowski2021survey,cuomo2022scientific}. Neural networks can autonomously learn artificial experiences and fit objective functions from high-dimensional parameter spaces, and significant progress has been made in knowledge-based physical problems during the last several years\cite{zhang2018application,gholami2020comparison}. Obiols-Sales O et al.\cite{obiols2020cfdnet} proposed that CFD-Net combines physical simulation and deep learning in a coupled framework to accelerate the convergence of Reynolds Averaged Navier-Stokes simulations. Neural networks are also trained to learn the lift coefficients of airfoils with various shapes under different flow Mach numbers, Reynolds numbers, and diverse angles of attack\cite{zhang2018application}. Gholami et al. \cite{gholami2020comparison} used neural networks along with numerical simulation to estimate flow depth variables and velocity for typical cases such as a 60° bend in a tube. The intelligent solution of PDEs has become increasingly prevalent and impactful in academic research. The PINNs designed by Raissi M\cite{raissi2019physics} et al. employ a DNN to approximate the solutions of PDEs, embedding the residuals that govern the PDE and its initial/boundary conditions within the loss function. Various variants of PINNs have been proposed to enhance their accuracy and efficiency. CanPINN\cite{chiu2022can} combines automatic and numerical differentiation in PINNs to improve training efficiency and accuracy. Meanwhile, ATLPINNs\cite{yan2023auxiliary} address the issues of low accuracy and non-convergence in original PINNs via auxiliary-task learning. These variants aim to strengthen PINN's performance in solving PDEs.

In addition to applying artificial intelligence in physics for solving PDEs, some researchers have explored incorporating artificial intelligence into mesh generation, mesh quality discrimination, and other areas to overcome the limitations of automation and intelligence in traditional mesh methods. Chen\cite{chen2021mve} presented a neural network-based mesh quality indicator for 3D cylinder meshes, along with a benchmark dataset. Wang\cite{wang2022evaluating} developed an algorithm to convert mesh data into graph data, subsequently introducing a deep graph neural network, GMeshNet, for mesh quality evaluation. MQENet\cite{zhang2023mqenet} is introduced as a structured mesh quality evaluation network that leverages dynamic graph attention, treating mesh evaluation as a graph classification task. Yilmaz and Kuzuoglu \cite{yilmaz2009particle} proposed a particle swarm optimization algorithm aimed at optimizing hexahedral mesh quality, addressing the inefficacies of the Laplace mesh smoothing algorithm in concave areas. Gargallo-Peiró A\cite{gargallo2014surface} suggested an optimization method for triangular and quadrilateral meshes on parameterized surfaces, focusing on node relocation to enhance mesh quality and prevent element inversion. Data-driven approaches to mesh generation have seen various implementations. Lowther et al.\cite{lowther1993density} applied neural networks to mesh density prediction in finite element mesh adaptation, using density information provided by the neural network to determine the size and placement of elements. MeshingNet \cite{zhang2020meshingnet} leverages posterior error estimates on an initial coarse mesh to predict non-uniform mesh densities for refinement. However, generating the training dataset is costly, and the model is targeted towards two-dimensional unstructured meshes. Huang et al. \cite{huang2021machine} developed a convolutional neural network to predict optimal mesh densities for diverse geometries. The validation phase required more than 60,000 simulations, supported by a training dataset of 20,000 simulations. Such extensive data demands may present challenges regarding computational resources and time allocation. Alexis Papagiannopoulos\cite{papagiannopoulos2021teach} introduced a neural network scheme for creating 2D simplicial meshes, using networks trained on meshed contour data to approximate vertex count, placement, and connectivity, though requiring extensive datasets for contours with a high number of vertices. Peng et al. \cite{peng2022automatic} employed two Artificial Neural Networks (ANNs) trained on an initial hybrid grid to predict point advancement and control grid sizing, with the ANN-based Advancing Layer Method generating initial training grids. Kim et al.\cite{kim2023gmr} utilized Graph Convolutional Neural Networks to learn local error densities and predict target mesh edge lengths, generating 16,000 control and feature meshes for training. However, these supervised learning meshing techniques necessitate significant effort in producing large training datasets, complicated by the costly creation and processing of these datasets, particularly for meshes with complex geometries or high vertex counts.
Moreover, generating training data demands domain-specific expertise. Physics-informed methods utilizing unsupervised learning to minimize the need for extensive training data in mesh generation have emerged. MGNet\cite{chen2022mgnet} proposed an intelligent approach for structured mesh creation, which is the first application of PINNs in this field. The network is designed to unsupervisedly learn meshing rules between parametric and computational domains, requiring only boundary specifications. Nevertheless, MGNet\cite{chen2022mgnet} faces challenges like training instability and extended durations. Chen \cite{chen2022improved} enhanced mesh quality by modifying loss terms with an auxiliary line strategy, initially requiring the manual selection of these lines as ground truth. However, it is still oriented to a 2D problem and requires additional manual intervention for drawing auxiliary lines. MeshNet first generates a coarse mesh using the traditional method and then adds the points of the coarse mesh as data items inside the training of the PINN, improving the quality of the generated mesh\cite{chen2023developing}. However, it introduces additional overhead for computation and training. While physics-informed methods alleviate the need for extensive manual data, they encounter issues such as slow training speeds, unstable convergence, and long training times. Furthermore, most of these approaches target 2D mesh generation, leaving the intelligent generation of 3D structured mesh remains an open problem.

\section{3DMeshNet: A Three-Dimensional Differential Neural Network for Structured Mesh Generation}
\label{method}
\subsection{Problem Setup}
The task of structured mesh generation involves the application of a potential mapping between parametric and computational domain to a given geometry. Typically, the parametric domain, represented by the coordinates $\left( \xi ,\eta ,\zeta \right) $, is defined as a basic cube [0×1]×[0×1]×[0×1], with each edge oriented either horizontally or vertically. The primary goal of any structured mesh generation technique is to create a mapping $\left( \xi ,\eta ,\zeta \right) $ to $\left( x,y,z \right) $ between these two spaces. This mapping should be unique and smooth, ensuring that the resulting mesh accurately conforms to the geometry's boundaries that require discretization.

In numerical computing, algebraic methods\cite{gordon1973construction} and PDE\cite{thompson1985numerical} methods are the two most commonly used structured mesh generation techniques. Algebraic methods use algebraic interpolation to describe the potential mapping relationship between the parametric domain $\left( \xi ,\eta ,\zeta \right) $ and the computational domain $\left( x,y,z \right) $.

The generation of meshes in 3D spaces through PDEs involves determining the correspondence between node coordinates in computational and parametric spaces. This is achieved by treating the process as a boundary-value problem solved by a system of elliptic PDEs. In this approach, elliptic equations are employed for 3D mesh generation. The process first solves for parametric coordinates within the parametric domain using elliptic mesh generation methods. Subsequently, these parametric meshes are transformed back into the computational domain, resulting in the acquisition of the computational surface mesh coordinates.

The governing differential equation is of the form of 
\begin{equation}
\nabla ^2\xi ^i=P^i\left( i=1,2,3 \right),
\end{equation}
or
\begin{equation}
\left\{ \begin{array}{c}\xi _{xx}+\xi _{yy}+\xi _{zz}=P\left( \xi ,\eta ,\zeta \right)\\	\eta _{xx}+\eta _{yy}+\eta _{zz}=Q\left( \xi ,\eta ,\zeta \right)\\	\zeta _{xx}+\zeta _{yy}+\zeta _{zz}=R\left( \xi ,\eta ,\zeta \right).\\\end{array} \right.
\end{equation}
where \(\nabla ^2\) denotes the Laplacian operator in Cartesian coordinate system,\(P^i = P^i\left( \xi ^1,\xi ^2,\xi ^3 \right)\), \(P^1=P\), \(P^2=Q\), and \(P^3=R\) are referred to as source terms. The source terms \(P\), \(Q\), and \(R\) are helpful in stretching the mesh surface \((\xi, \eta, \zeta)\) respectively. Negative values of the source terms cause the corresponding mesh faces to move toward the decreasing curve coordinates, while positive values of the source terms have the opposite effect. Therefore, the mesh density can be adjusted by adjusting the values of the source terms.

\subsection{The Architecture of 3DMeshNet}
We propose 3DMeshNet, a novel neural network-based approach for structured 3D mesh generation. It leverages the representation learning capabilities of neural networks to learn the rules governing mesh partitioning and determine the mapping between parametric and computational domain by encoding point coordinates through a deep network. Specifically, the network mesh partitioning rules by minimizing a weighted loss function comprising the residuals of governing differential equations and surface constraint. Once trained, 3DMeshNet can generate meshes of arbitrarily specified sizes for the corresponding geometries in a feedforward.

\begin{figure*}[!t]
\centering

\includegraphics[width=16cm]{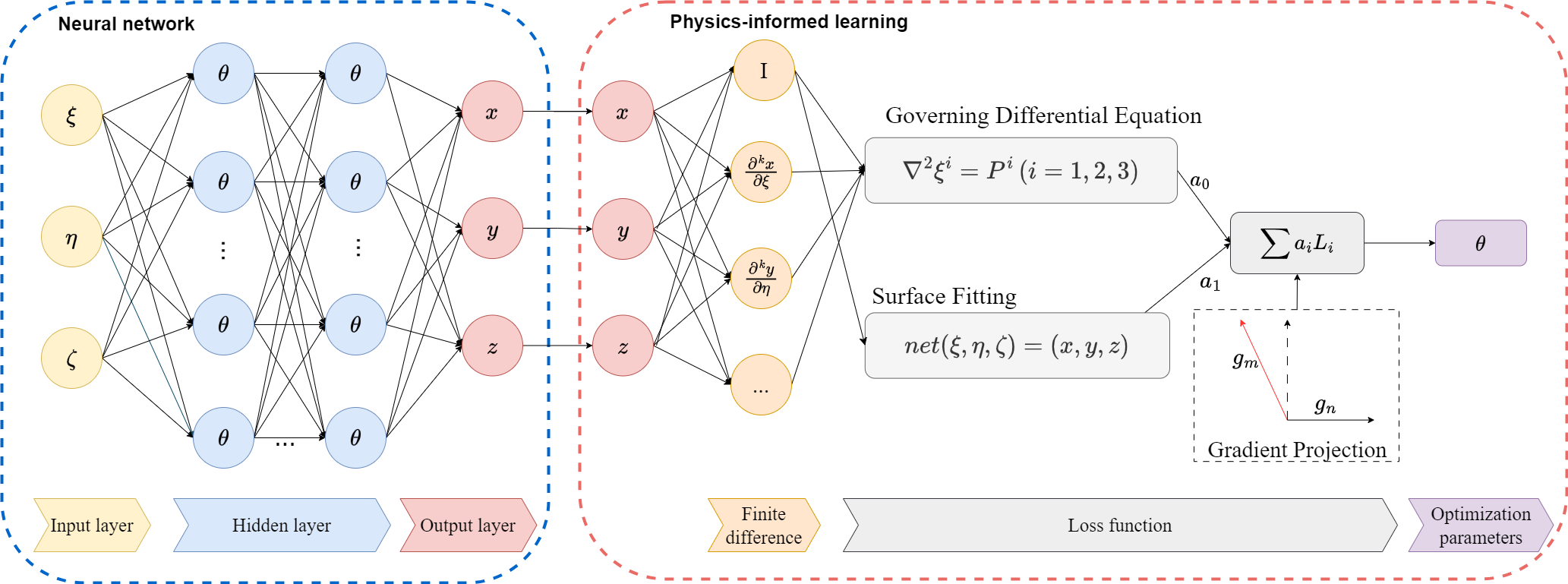}

\caption{The architecture of the proposed 3DMeshNet. It fuses a Neural network, which learns mesh partitioning from 3D points, with a Physics-informed learning part that enforces physical laws via elliptic differential equations. The Neural network extracts features and predicts mesh coordinates, while the Physics-informed learning part uses a finite difference layer for efficient derivatives calculation, guiding the NN's training.}
\label{fig_3DMeshNet}
\end{figure*}

The architecture of 3DMeshNet is illustrated in Fig.~\ref{fig_3DMeshNet}. It comprises a Neural network and a Physics-informed learning part. The Neural network component primarily learns the rules of mesh partitioning, while the Physics-informed learning part provides physical information from the elliptic governing differential equations. 

The input to 3DMeshNet is sampled points from the interior and surface points of the parametric domain. Each input comprises a set of 3D points$=\left\{ \left( \xi _i,\eta _i,\zeta _i \right) \right\} _{i=1}^{i=N}$, where N is the number of points. The network output consists of points $\left\{ \left( x_i,y_i,z_i \right) \right\} _{i=1}^{i=N}$ representing the mesh in the computational domain. 

The Neural network portion contains an input, multiple hidden, and output layers. The input layer takes training points $(\xi ,\eta ,\zeta )$ and passes them to hidden layers. Within hidden layers, the network performs computations to extract high-dimensional features via Eq.~\ref{eq:hidden}.:

\begin{equation}
\chi^k=\sigma \left( \chi^{k-1}W^{k-1}+b^{k-1} \right) ,
\label{eq:hidden}
\end{equation}
where $\chi$ denotes a hidden layer output, \(\sigma\) the activation function, and $W$ and $b$ the layer weights and bias, respectively. Subsequently, the output layer produces computational domain coordinate predictions based on received hidden representations. 

In the Physics-informed learning part, elliptic control equations are employed to guide the Neural network's training in learning mesh partitioning rules, with boundary conditions describing the surface fitting of the given 3D object. In the Finite Difference layer, 3DMeshNet applies the Finite Difference method, based on the Taylor series expansion, to improve efficiency in derivative computation. The loss function is determined through Loss Function Reweighting, informed by the 3D elliptical equation and Surface Fitting. Loss Function Reweighting employs a multi-task learning strategy and surface point weighting scheme to reweight the two-part loss. The multi-task learning strategy leverages homoscedastic uncertainty principles from Bayesian inference to adjust weights $a$ according to task-specific uncertainties. The surface point weighting scheme introduces a surface point weighting scheme, assigning weights based on the Euclidean distance between predicted and actual surface points, thereby enhancing boundary mesh quality. Moreover, gradient projection effectively mitigates gradient conflict between the elliptic equation gradient and the Surface Fitting gradient. Lastly, network parameters are then updated across training sessions through loss backpropagation.
\subsection{Loss Function with Finite Difference}
\label{FDMethod}
The loss function of 3DMeshNet consists of two parts: a boundary loss for surface fitting and a 3D elliptic PDE system as basic governing differential equations. The 3D elliptic PDE provides the physical-informed learning aspect for mesh generation. By embedding the 3D elliptic PDE into the loss function, we aim to minimize it during the unsupervised learning process.
\begin{equation}
\begin{split}
\alpha_1 x_{\xi \xi} + \alpha_2 x_{\eta \eta} + \alpha_3 x_{\zeta \zeta} &+ 2\beta_{12} x_{\xi \eta} + 2\beta_{23} x_{\eta \zeta} + 2\beta_{31} x_{\zeta \xi} = 0 \\
\alpha_1 y_{\xi \xi} + \alpha_2 y_{\eta \eta} + \alpha_3 y_{\zeta \zeta} &+ 2\beta_{12} y_{\xi \eta} + 2\beta_{23} y_{\eta \zeta} + 2\beta_{31} y_{\zeta \xi}= 0 \\
\alpha_1 z_{\xi \xi} + \alpha_2 z_{\eta \eta} + \alpha_3 z_{\zeta \zeta} &+ 2\beta_{12} z_{\xi \eta} + 2\beta_{23} z_{\eta \zeta} + 2\beta_{31} z_{\zeta \xi}= 0
\end{split}
\end{equation}
\begin{equation}
\begin{split}
\alpha_1 &= (x_{\eta}^2 + y_{\eta}^2 + z_{\eta}^2)(x_{\zeta}^2 + y_{\zeta}^2 + z_{\zeta}^2) - (x_{\eta}x_{\zeta} + y_{\eta}y_{\zeta} + z_{\eta}z_{\zeta}), \\
\alpha_2 &= (x_{\zeta}^2 + y_{\zeta}^2 + z_{\zeta}^2)(x_{\xi}^2 + y_{\xi}^2 + z_{\xi}^2) - (x_{\xi}x_{\zeta} + y_{\xi}y_{\zeta} + z_{\xi}z_{\zeta}), \\
\alpha_3 &= (x_{\xi}^2 + y_{\xi}^2 + z_{\xi}^2)(x_{\eta}^2 + y_{\eta}^2 + z_{\eta}^2) - (x_{\eta}x_{\xi} + y_{\eta}y_{\xi} + z_{\eta}z_{\xi}), \\
\beta_{12} &= (x_{\eta}x_{\zeta} + y_{\eta}y_{\zeta} + z_{\eta}z_{\zeta})(x_{\zeta}x_{\xi} + y_{\zeta}y_{\xi} + z_{\zeta}z_{\xi}) \\
&\quad - (x_{\xi}x_{\eta} + y_{\xi}y_{\eta} + z_{\xi}z_{\eta})(x_{\zeta}^2 + y_{\zeta}^2 + z_{\zeta}^2), \\
\beta_{23} &= (x_{\xi}x_{\zeta} + y_{\xi}y_{\zeta} + z_{\xi}z_{\zeta})(x_{\eta}x_{\xi} + y_{\eta}y_{\xi} + z_{\eta}z_{\xi}) \\
&\quad - (x_{\zeta}x_{\eta} + y_{\zeta}y_{\eta} + z_{\zeta}z_{\eta})(x_{\xi}^2 + y_{\xi}^2 + z_{\xi}^2), \\
\beta_{31} &= (x_{\xi}x_{\eta} + y_{\xi}y_{\eta} + z_{\xi}z_{\eta})(x_{\eta}x_{\zeta} + y_{\eta}y_{\zeta} + z_{\eta}z_{\zeta}) \\
&\quad - (x_{\zeta}x_{\xi} + y_{\zeta}y_{\xi} + z_{\zeta}z_{\xi})(x_{\eta}^2 + y_{\eta}^2 + z_{\eta}^2).
\end{split}
\end{equation}
where $x_{\xi}$ represents the first-order partial derivative of $x$ with respect to$\xi $, while $x_{\xi \eta}$ signifies the second-order partial derivative of $x$, taken sequentially first with respect to $\xi $ and then $\eta $. This notation is similarly extended to other second-order derivatives. Then, the form of the loss function with governing differential equations loss and geometric surface constraint is as follows:
\begin{equation}
\begin{split}
Loss\left( x,y,z \right) =Loss_1+Loss_2,
\end{split}
\end{equation}
where,
\begin{equation}
\begin{split}
\label{loss1_loss2}
Loss_1 &= \frac{1}{N_r}\sum_{n=1}^{N_r} \bigg( |\alpha _1x_{\xi \xi} + \alpha _2{x_{\eta}}_{\eta} + \alpha _3x_{\zeta \zeta} \\
&\quad + 2\beta _{12}x_{\xi \eta} + 2\beta _{23}x_{\eta \zeta} + 2\beta _{31}x_{\zeta \xi}|^2 \\
&\quad + |\alpha _1y_{\xi \xi} + \alpha _2{y_{\eta}}_{\eta} + \alpha _3y_{\zeta \zeta} \\
&\quad + 2\beta _{12}y_{\xi \eta} + 2\beta _{23}y_{\eta \zeta} + 2\beta _{31}y_{\zeta \xi}|^2 \\
&\quad + |\alpha _1z_{\xi \xi} + \alpha _2{z_{\eta}}_{\eta} + \alpha _3z_{\zeta \zeta} \\
&\quad + 2\beta _{12}z_{\xi \eta} + 2\beta _{23}z_{\eta \zeta} + 2\beta _{31}z_{\zeta \xi}|^2 \bigg), \\
Loss_2 &= \frac{1}{N_b}\sum_{n=1}^{N_b} \bigg( |(x_n,y_n,z_n) - f(\xi _{n,}\eta _n,\zeta _n)|^2 \bigg).
\end{split}
\end{equation}
the loss function is composed of two parts: the first part, $Loss_1$ is the control equation loss term, and the second part, $Loss_2$ is the geometric surface constraint term. Here, $N_r$ and $N_b$ represent the number of training points for the control equation and the geometric surface. $\left( x_n,y_n,z_n \right) $ denotes the coordinates of the nth actual surface point, and $f(\xi _{n,}\eta _n,\zeta _n)$ refers to the predicted value corresponding to the input at the surface point.

To achieve a more efficient implementation, we do not use automatic differentiation for derivative calculations because it is time-consuming \cite{chiu2022can}. Instead, we shift to using the finite differences method for calculating derivatives. The finite differences method is based on the Taylor series expansion, allowing us to compute derivatives more efficiently and effectively.

\begin{figure}[!t]
\centering
\includegraphics[width=1.5in]{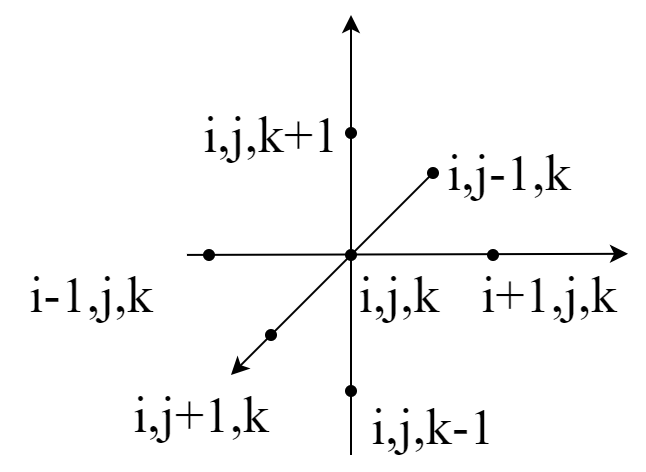}
\caption{Coordinate points of the three-dimensional finite difference.}
\label{FD}
\end{figure}

As shown in Fig.~\ref{FD}, the central point \(i\), \(j\) ,\(k\) serves as the reference point, with its neighboring points in the three orthogonal directions labeled as$\mathrm{ (}i+1,j,k$), ($i-1,j,k$), $(i,j+1,k)$, ($i,j-1,k$), ($i,j,k+1$), and ($i,j,k-1)$. The relative positions of these points to the central point represent the layout of points in the finite differences framework used for calculating partial derivatives. By combining the function values at these points with the function value at the central point and applying the Taylor expansion, one can approximate the partial derivatives at the central point.

Specifically, the derivation formula for the finite differences method is as follows:
\begin{equation}
\begin{split}
\left\{ \begin{array}{l}	\vec{r}_{\xi}=\left( \vec{r}_{i+1,j,k}-\vec{r}_{i-1,j,k} \right) /2h_1\\	\vec{r}_{\eta}=\left( \vec{r}_{i,j+1,k}-\vec{r}_{i,j-1,k} \right) /2h_2\\	\vec{r}_{\zeta}=\left( \vec{r}_{i,j,k+1}-\vec{r}_{i,j,k-1} \right) /2h_3,\\\end{array} \right.
\end{split}
\end{equation}

\begin{equation}
\begin{split}
\begin{cases}	\vec{r}_{\xi \xi}=\left( \vec{r}_{i+1,j,k}-2\vec{r}_{i,j,k}+\vec{r}_{i-1,j,k} \right) /h_{1}^{2}\\	\vec{r}_{\eta \eta}=\left( \vec{r}_{i,j+1,k}-2\vec{r}_{i,j,k}+\vec{r}_{i,j-1,k} \right) /h_{2}^{2}\\	\vec{r}_{\zeta \zeta}=\left( \vec{r}_{i,j,k+1}-2\vec{r}_{i,j,k}+\vec{r}_{i,j,k-1} \right) /h_{3}^{2},\\\end{cases}
\end{split}
\end{equation}

\begin{equation}
\begin{cases}
	\begin{aligned}
		\vec{r}_{\xi \eta} &= \frac{1}{4h_1h_2}\left( \vec{r}_{i+1,j+1,k} - \vec{r}_{i+1,j-1,k} \right. \\
		&\quad \left. - \vec{r}_{i-1,j+1,k} + \vec{r}_{i-1,j-1,k} \right), \\
		\vec{r}_{\eta \zeta} &= \frac{1}{4h_2h_3}\left( \vec{r}_{i,j+1,k+1} - \vec{r}_{i,j+1,k-1} \right. \\
		&\quad \left. - \vec{r}_{i,j-1,k+1} + \vec{r}_{i,j-1,k-1} \right), \\
		\vec{r}_{\zeta \xi} &= \frac{1}{4h_3h_1}\left( \vec{r}_{i+1,j,k+1} - \vec{r}_{i+1,j,k-1} \right. \\
		&\quad \left. - \vec{r}_{i-1,j,k+1} + \vec{r}_{i-1,j,k-1} \right),
	\end{aligned}
\end{cases}
\end{equation}
where $\vec{r}$ denotes the$(x,y,z)$ vector and $h_1$ , $h_2$ , $h_3$ represent the steps in the three spatial directions.

\subsection{Loss Function Reweighting}
\label{LossFunctionReweighting}

As the loss function consists of two components, namely the Governing Differential Equation and the Surface Fitting Loss, we can consider these two optimization objectives as a multi-task optimization problem. Moreover, a significant numerical difference exists between the losses of these two parts. Weight hyper-parameters are expensive to tune and often take time for each trial. Therefore, it is desirable to find a more convenient approach that can learn the optimal weights. Consequently, We incorporate a strategy for multi-task learning as presented in source\cite{kendall2018multi} to assign weight appropriately to the losses of distinct tasks. 

The methodology utilizes homoscedastic uncertainty principles from Bayesian inference to assess and articulate the variance in confidence across tasks, enabling a sophisticated approach to loss weighting based on task-specific uncertainties. Let $NN(X;\theta)$ denote the output of a neural network with input $x$ and $u$ is the model's output, in this paper represented as (x, y, z). We define the following probabilistic model:
\begin{equation}
\begin{split}
\label{probability distribution}
p\left( u |NN\left( X;\theta \right) \right) =\mathcal{N} \left( NN\left( X;\theta \right) ,a^2 \right),
\end{split}
\end{equation}
and Eq.\ref{probability distribution} describes the conditional probability distribution of the target variable \( u \) given the input \( X \) and the neural network prediction \( NN(X;\theta) \). We assumes that the distribution of the target variable \( u \) follows a Gaussian distribution with mean equal to the predicted value \( NN(X;\theta) \) and variance \( a^2 \).With the scalar $a$ as a noise parameter. In a setting of a neural network with K tasks, the multi-objective likelihood can be written as:
\begin{align}
&p(u_1, \ldots, u_K | NN(X;\theta)) \nonumber \\
&= p(u_1 | NN(X;\theta)) \cdots  p(u_K | NN(X;\theta)).
\end{align}

In maximum likelihood estimation, it is often convenient to work with the natural logarithm of the likelihood function, the log likelihood can be written as:
\begin{equation}
\begin{aligned}
&\log p(u|NN(X;\theta)) \\
&= \log \left[ \left( 2\pi a^2 \right) ^{-\frac{1}{2}}\exp \left( -\frac{1}{2a^2}\left\| u-NN\left( X;\theta \right) \right\| ^2 \right) \right] \\
&= \log \left( 2\pi a^2 \right) ^{-\frac{1}{2}} + \log \left[ \exp \left( -\frac{1}{2a^2}\left\| u-NN\left( X;\theta \right) \right\| ^2 \right) \right] \\
&= -\frac{1}{2}\log \left( 2\pi a^2 \right) - \frac{1}{2a^2}\left\| u-NN\left( X;\theta \right) \right\| ^2 \\
&= -\frac{1}{2}\log \left( 2\pi \right) - \frac{1}{2}\log \left( a^2 \right) - \frac{1}{2a^2}\left\| u-NN\left( X;\theta \right) \right\| ^2,
\end{aligned}
\end{equation}

\begin{equation}
  \log p(u|NN(X;\theta))\propto -\frac{1}{2a^2}||u-NN(X;\theta)||^2-\log a , 
\end{equation}
and we can therefore maximize:

\begin{equation}
\begin{aligned}
\label{logp}
&\log p\left( u_1,...,u_K\mid NN(X;\theta ) \right) \\
    &=\log p\left( u_1\mid NN(X;\theta ) \right) +...+\log p\left( u_K\mid NN(X;\theta ) \right) \\
    &=\log \mathcal{N} \left( u_1;NN(X;\theta ),a_{1}^{2} \right) +...+\log \mathcal{N} \left( u_K;NN(X;\theta ),a_{K}^{2} \right) \\
    &=(-\frac{1}{2\alpha _{1}^{2}}L_1-\log a_1)+...+(-\frac{1}{2\alpha _{K}^{2}}L_K-\log a_K) \\
    &=(-\frac{1}{2\alpha _{1}^{2}}L_1-\frac{1}{2}\log a_{1}^{2})+...+(-\frac{1}{2\alpha _{K}^{2}}L_K-\frac{1}{2}\log a_{K}^{2}) \\
    &\propto (-\frac{1}{2\alpha _{1}^{2}}L_1-\log\mathrm{(}1+a_{1}^{2}))+...+(-\frac{1}{2\alpha _{K}^{2}}L_K-\log\mathrm{(}1+a_{K}^{2})).
\end{aligned} 
\end{equation}

According to Eq.~\ref{logp}, to facilitate optimization, we prefer to maximize the sum of the logarithms of these likelihoods, which simplifies to a sum of terms involving task-specific losses $L_i$ and noise parameters $a _i$. The logarithm of $a^2$ is modified to $\log(1+a^2)$, ensuring the expression remains positive and avoids training issues where the loss might otherwise become negative. This leads to our minimization target:
\begin{equation}
\begin{split}
L\left( \theta ,a _1,...,a _K \right) =\sum_i^K{\left( \frac{1}{2a _{i}^{2}}L_i+\log \left( 1+a _{i}^{2} \right) \right)},
\end{split}
\end{equation}
this objective function balances each task's loss against its uncertainty, represented by $a$, now a parameter to be learned alongside the neural network's weights. It allows adaptive learning of each task's importance, preventing the network from discounting easier tasks with smaller uncertainties. Additionally, the term \(\log(1+a_{i}^{2})\)
 acts as a regularization, averting excessively large values of $a$ and ensuring a balanced task weighting within the multi-task learning framework.

Another innovation in Loss Function Reweighting is the implementation of a surface point weighting scheme. This scheme emphasizes the importance of accurately capturing surface conditions by assigning higher weights to poorly predicted surface points and lower weights to well-predicted ones. As a result, 3DMeshNet can more effectively learn and conform to surface conditions, enhancing the mesh quality at the boundaries.
\begin{equation}
weight=\left\| (x^{true}-x^{pre},y^{true}-y^{pre},z^{true}-z^{pre}) \right\| +1,
\end{equation}
where $\left( {x}^{pre},{y}^{pre},{z}^{pre} \right) $ is the predicted output of the ith surface point and $\left( {x}^{true},{y}^{true},{z}^{true} \right) $ is the coordinates of the ith real surface point. After calculating the Euclidean distance between the predicted surface point and the real surface point, the distance is added by $1$ to get the weight $w=distance+1$, or directly use the distance as the weight $w=distance$.

After introducing the loss function adaptive weighting strategy as well as the surface point weighting mechanism, the loss function of Eq.~\ref{lassall} can be rewritten in the following form

\begin{equation}
\begin{aligned}
Loss\left( x,y,z \right) =\frac{1}{2a_{1}^{2}}Loss_1+\frac{1}{2a_{2}^{2}}weight\cdot Loss_2
\\
+\log \left( 1+a_{1}^{2}+a_{2}^{2} \right). 
\label{lassall}
\end{aligned}
\end{equation}
where, $Loss_1$ and $Loss_2$ are equal to the loss of Equation \ref{loss1_loss2}. 3DMeshNet finally uses the Eq.~\ref{lassall} as a loss function to guide the training of the network.

\subsection{Gradient Projection}
\label{Gradient Projection}

In addition to balancing the Surface Fitting loss with the Governing Differential Equation loss efficiently, our approach introduces gradient projection to address the issue of ill-conditioned gradients from a gradient perspective. Previous works, such as MGNet\cite{chen2022mgnet} and MeshNet \cite{zhang2020meshingnet}, have focused primarily on numerical improvements to the loss function, with little exploration into enhancements from a gradient standpoint. Furthermore, upon examining the loss function, it is evident that substantial oscillations in the loss are not conducive to the optimal collection of structured mesh results. Therefore, we have incorporated the gradient projection mechanism to improve network convergence. This novel methodology offers a comprehensive solution by tackling numerical and gradient aspects, enhancing the overall efficacy of structured mesh generation.

We define $\varphi _{mn}$ as the angle between two task gradients $g_m$ and $g_n$, and when $\cos \varphi _{mn}<0$ it means that the gradients are conflicting. Specifically, mitigating the gradient conflict of the loss function can include the following steps:

(1)	Assume that the gradient of task $T_m$ is $g_m$ and the gradient of task $T_n$ is $g_n$.

(2)	Determine whether $g_m$ conflicts with $g_n$ by calculating the cosine similarity between the vectors $g_m$ and $g_n$, where a negative value indicates the gradient of the conflict. The cosine similarity is calculated as follows.

\begin{equation}
\begin{aligned}
\cos \varphi _{mn}=\frac{g_m\cdot g_n}{\left\| g_m \right\| \times \left\| g_n \right\|}.
\end{aligned}
\end{equation}

(3)	If the cosine similarity is negative, replace $g_n$ with its projection $g_{n}^{PC}$ on the normal plane of $g_n$, that is, $g_{m}^{PC}=g_{m}^{PC}-\frac{g_{m}^{PC}\cdot g_n}{\left\| g_n \right\| ^2}g_n$. If the gradients are not conflicting, $\cos \varphi _{mn}>0$ is non-negative, and the original gradient $g_m$ is kept constant.

\begin{figure*}[htbp]
	\centering
	\captionsetup[subfloat]{font=scriptsize} 
	\subfloat[No conflict in task gradients]{\includegraphics[width=.35\columnwidth]{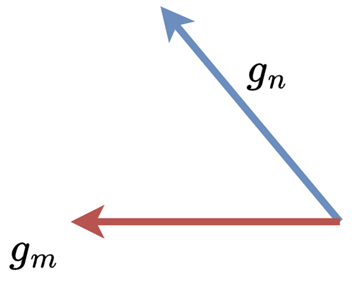}}\hspace{14pt}
	\subfloat[Conflicts in task gradients]{\includegraphics[width=.35\columnwidth]{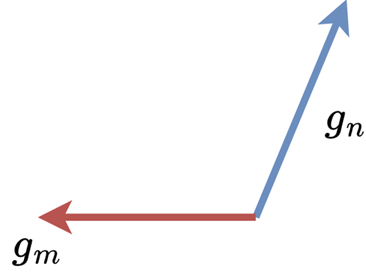}}\hspace{14pt}
	\subfloat[Projects task $n$’s gradient onto the normal vector of task $m$’s gradient]{\includegraphics[width=.35\columnwidth]{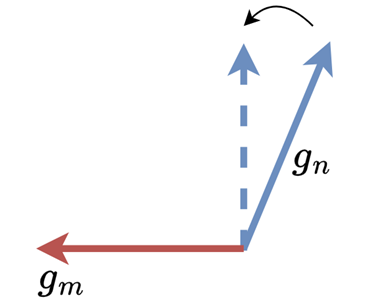}}
	\caption{Conflicting gradients and gradient projections.}
 \label{pg-all}
\end{figure*}

Fig~\ref{pg-all}. shows the principle of gradient conflict resolution. (a) shows the gradient projection algorithm does not change the gradients of non-conflicting tasks. (b) Tasks $m$ and $n$ have conflicting gradient directions. The gradient projection algorithm in the case of conflicting gradients is shown in (c), which projects the gradient of task $n$ onto the normal vector of the gradient of task $m$.

\section{Experimental results}
\label{Experimental}
In this section, we demonstrate the effectiveness of the proposed 3DMeshNet in both 2D and 3D mesh generation cases. In 2D-case1 and 2D-case2, we use a configuration with hidden layer sizes of \{45,45,45\}; for 3D-case1, 3D-case2, and 3D-case3, we use a configuration with \{85,85,85,85\} hidden layer sizes; and for 3D-case4, we use a configuration with \{128,128,128,128\} MLP. Tanh serves as the activation function in all cases. Due to data dimension limitations, we opt for a lightweight network model to extract features instead of employing larger-scale models. 3DMeshNet randomly samples 4000 internal points and 6000 surface points from the parametric domain for training during each training iteration, and it employs Adam and L-BFGS-B optimizers to refine network parameters. Specifically, 3DMeshNet first employs the Adam optimizer for 12000 iterations of training. The learning rate is 1e-3, and it is reduced by a factor of 0.9 every 1000 epochs. Subsequently, the L-BFGS-B is employed to optimize the model parameters further. All experiments are implemented using PyTorch. Although the training processes are conducted on the 3080Ti GPU, due to the lightweight network model, the meshing process can be performed rapidly on CPU platforms. To validate the effectiveness of 3DMeshNet, we compared it with TFI\cite{gordon1973construction} and PDE\cite{thompson1985numerical}. As NN-based models, we also selected PINN\cite{raissi2019physics} and MGNet\cite{chen2022mgnet} as backbones to compare mesh quality with 3DMeshNet, demonstrating its superiority in mesh generation.

In assessing the generated mesh quality, we employe the loss on the test set as an auxiliary indicator of the mesh quality. Additionally, we utilize a comprehensive evaluation using the following five criteria for mesh quality to determine the overall integrity of the generated mesh:

 \textbf{(1) Minimum Included Angle and Maximum Included Angle(2D, 3D):} Both Minimum Included Angle and Maximum Included Angle are measures of mesh skewness. For all types of meshes, Minimum Included Angle denotes the minimum angle of the mesh and Maximum Included Angle denotes the maximum angle of the mesh;

\textbf{(2) Equiangle Skewness(2D, 3D):} Equiangle Skewness is expressed as the maximum ratio of the angle of the mesh to the angle of the equiangular mesh. The angle skewness varies between 0 (good) and 1 (poor).

\textbf{(3) Aspect Ratio(2D, 3D):} The hexahedral cell aspect ratio is computed from the ratio of the maximum length, width, and height and the minimum length, width, and height. The aspect ratio is always greater than or equal to 1, with a value of 1 representing a cube.

\textbf{(4) Centroid Skewness(3D):} Centroid Skewness is one minus the minimum dot product between the cell face normal and the vector connecting the cell centroid and the face centroid. Values range from 0 (no-skew) to 1 (collapsed cell). This measure is only valid for block cells.

\begin{figure}[!t]
\centering
\includegraphics[width=2.5in]{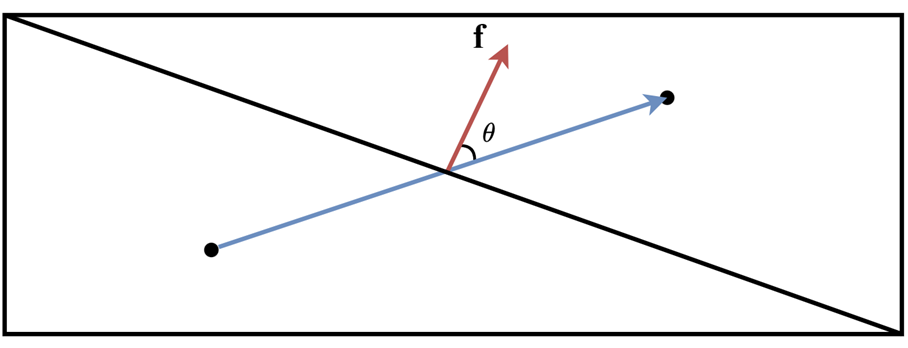}
\caption{Cell Non-Orthogonality; The concept of cell non-orthogonality, where the black dots represent the centroids of the cells, and $\boldsymbol{f}$ is the normal of the shared face. The angle $\boldsymbol{\theta }$ is defined by the vector normal to the shared face and the line connecting the two centroids.}
\label{CellNon}
\end{figure}

\textbf{(5) Cell Non-Orthogonality(3D):} This measure operates on pairs of neighboring cells that share a face. An angle is calculated between a line connecting the two cells' centroids and the shared face's normal. The maximum value of this measure for all faces of a given cell is reported per cell. The angle shown as $\theta $ in Fig.~\ref{CellNon}.

\textbf{(6) Meshing Overhead(2D, 3D):}
 This metric gauges the time required to generate a mesh on a given geometry.Use seconds (s) as the measurement unit.

 \textbf{(7) Mesh validity(2D, 3D):} This metric assesses the efficacy of the mesh generated on the geometry using the respective method. Meshes are deemed ineffective if there are notable instances of elements extending beyond boundaries or lying below the boundary shape. Numerically, in this paper, the number of cells with negative volume is considered to be within 0.5\% of the total number of cells as a valid mesh. Valid meshes are denoted by Yes (Y), and invalid meshes by No (N).

\subsection{Experimental results of 3DMeshNet in 2D structured mesh generation}

\begin{figure*}[htbp]
    \centering
    \captionsetup[subfloat]{font=scriptsize}
    \subfloat[TFI]{\includegraphics[width=.4\columnwidth]{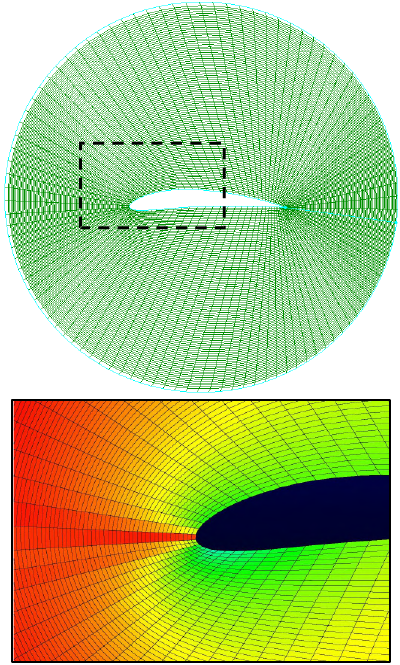}}\hspace{10pt}
    \subfloat[PDE]{\includegraphics[width=.4\columnwidth]{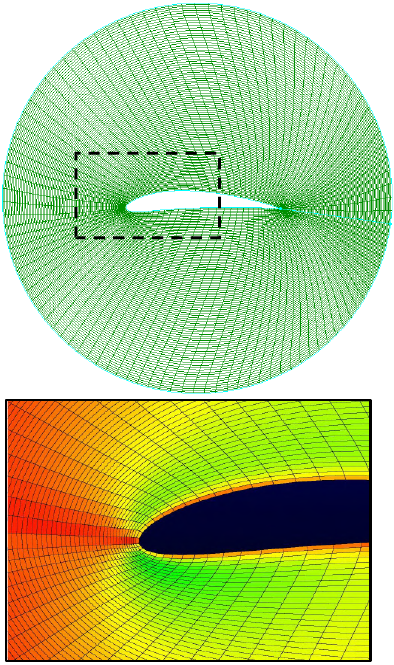}}\\
    \resizebox{0.35\columnwidth}{!}{\subfloat[PINN]{\includegraphics{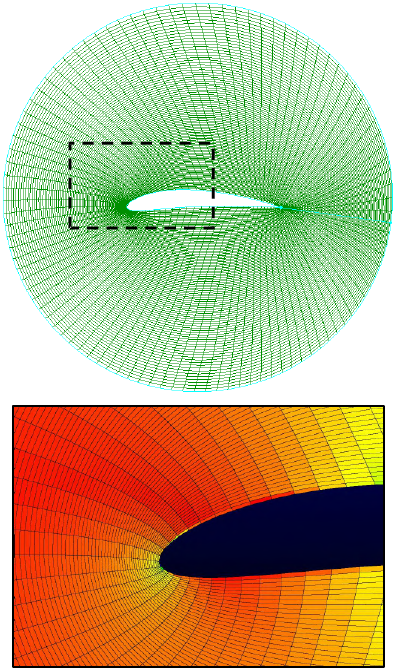}}}\hspace{15pt}
    \resizebox{0.35\columnwidth}{!}{\subfloat[MGNet]{\includegraphics{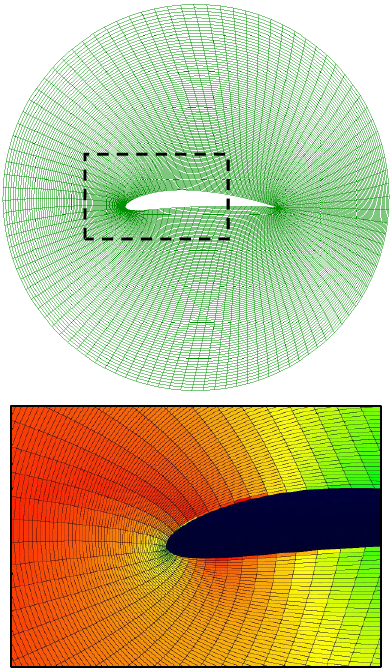}}}\hspace{15pt}
    \resizebox{0.35\columnwidth}{!}{\subfloat[3DMeshNet]{\includegraphics{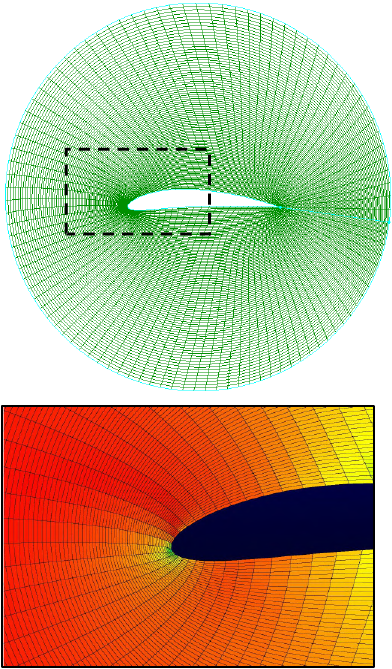}}} 
    \caption{Comparison of structured mesh generation results for 2D-case1 geometry using TFI, PDE, PINN, MGNet, and 3DMeshNet methods. The colors in the subfigures below indicate mesh quality: red represents higher orthogonality, and blue denotes lower.}
    \label{2d-1}
\end{figure*}

\begin{table*}[!t]
\centering
\caption{Numerical results comparison of structured mesh generation on 2D-case1 geometry using TFI, PDE, PINN, MGNet, and 3DMeshNet methods.}
\label{2d-1-t}
\begin{tabular}{lllllll}
\hline
Method & Average test set loss & Max. Included Angle & Min. Included Angle & Equiangle Skewness & Aspect Ratio & Mesh  Validity \\
\hline

TFI & - & 104.53220 & 75.61654 & 0.16243 & 4.91350 &Y\\ 
PDE & - & 105.50302 & 74.72664 & 0.17303 & 4.29008 &Y\\ 
PINN & \(5.0700 \times 10^{-6}\) & 104.10900 &75.86799 & 0.16059 & 4.52362 &Y\\ 
MGNet& \(9.5063\times 10^{-7}\) & 104.15644 & 75.99211 &0.15853 & \textbf{4.24366}&Y\\ 
3DMeshNet & \(\bm{3.4100 \times 10^{-7}}\) & \textbf{103.61675} & \textbf{76.31713} & \textbf{0.15812} & 4.24556 &Y\\ 
\hline
\end{tabular}
\end{table*} 

Table \ref{2d-1-t} and Fig.~\ref{2d-1} show the meshing results for 2D test case 1, where 3DMeshNet surpasses PINN and MGNet by achieving a lower loss. It showcases a maximum angle of 103.61 compared to PINN's and MGNet's 104.1 and a minimum angle of 76.31 against PINN's 75.86, indicating improved orthogonality. Compared with TFI and PDE methods, 3DMeshNet yields higher-quality meshes, outperforming PDE's 105.50 and 74.72. Accordingly, 3DMeshNet demonstrates superior orthogonality to both PDE and TFI methods in this test case.

\begin{figure*}[htbp]
    \centering
    \captionsetup[subfloat]{font=scriptsize}
    \subfloat[TFI]{\includegraphics[width=.4\columnwidth]{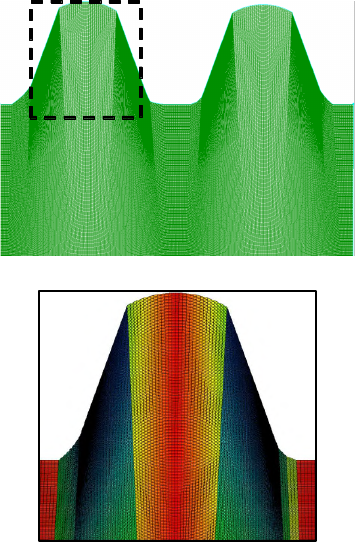}}\hspace{10pt}
    \subfloat[PDE]{\includegraphics[width=.4\columnwidth]{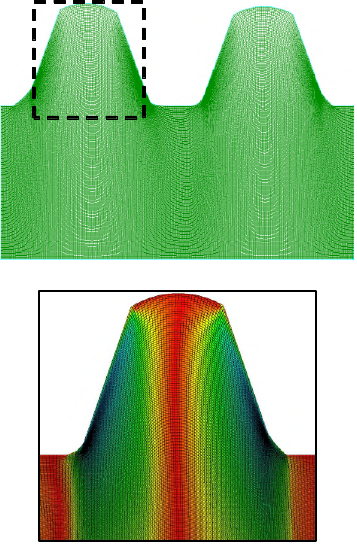}}\\
    \subfloat[PINN]{\includegraphics[width=.35\columnwidth]{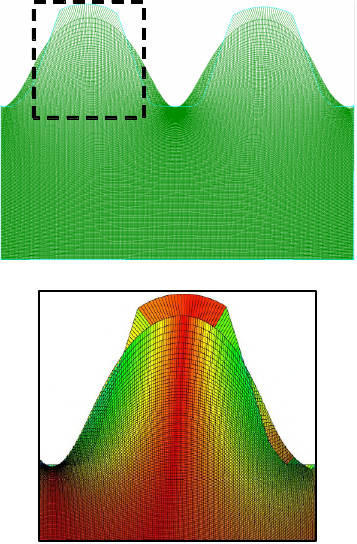}}\hspace{15pt}
    \subfloat[MGNet]{\includegraphics[width=.35\columnwidth]{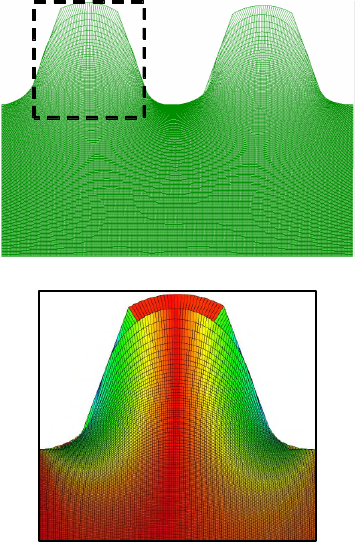}}\hspace{15pt}
    \subfloat[3DMeshNet]{\includegraphics[width=.35\columnwidth]{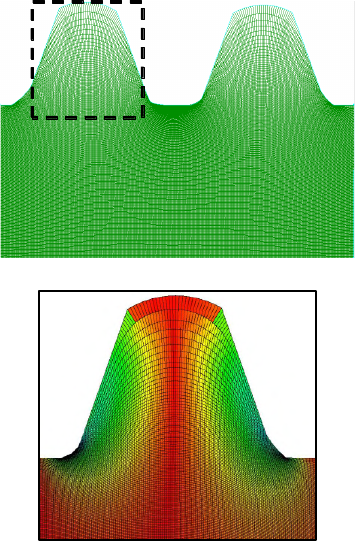}}    
    \caption{Comparison of structured mesh generation results for 2D-case2 geometry using TFI, PDE, PINN, MGNet, and 3DMeshNet methods. The colors in the subfigures below indicate mesh quality: red represents higher orthogonality, and blue denotes lower.}
    \label{2d-2}
\end{figure*}

\begin{table*}[htbp]
\centering
\caption{Numerical results comparison of structured mesh generation on 2d-case2 geometry using PINN, TFI, PDE, 3DMeshNet, and MGNet methods}
\label{2d-2-t}
\begin{tabular}{lllllll}
\hline
Method & Average Test Set Loss & Max. Included Angle & Min. Included Angle & Equiangle Skewness & Aspect Ratio & Mesh Validity\\ 
\hline
TFI & - & 118.6771 & 61.38721 & 0.31882 & 2.26686 & Y\\ 
PDE & - & 105.2497 & 74.80033 & 0.16962 & 2.03112 & Y\\
PINN & 0.00400 & \textbf{97.87325}&\textbf{82.13464} & \textbf{0.08773} & \textbf{1.05901} & N\\
MGNet & 0.00398 & 98.30392 & 81.73544 & 0.09252 & 2.06594 & Y\\ 
3DMeshNet & \textbf{0.00040} & 98.32237 & 81.72149 & 0.09268 & 2.07669 & Y\\ 

\hline
\end{tabular}
\end{table*}

Table~\ref{2d-2-t} and Fig.~\ref{2d-2} illustrate the results of structured mesh generation on 2D-case1 geometry using TFI, PDE, PINN, MGNet and 3DMeshNet methods. 3DMeshNet's average test set loss 0.0004 exceeds PINN's 0.0040 and MGNet's 0.003. In terms of mesh quality, 3DMeshNet has a Maximum Included Angle of 98.32 compared to PINN's 97.87 and a Minimum Included Angle of 81.72 compared to PINN's 82.13, and there appears to be no numerical advantage for 3DMeshNet. However, PINN's meshes displayed negative areas and poor boundary fitting capabilities, indicating a failure to learn the shape information accurately, and Mesh Validity is False. Moreover, 3DMeshNet exhibits a lower  Skewness Equiangle of 0.09 compared to TFI's 0.31 and PDE's 0.16, indicating a lesser degree of mesh skewness.

\subsection{Experimental results of 3DMeshNet in 3D structured mesh generation}
In this section, we perform experiments on four shapes to demonstrate the efficacy of 3DMeshNet in generating 3D structured meshes. To evaluate the quality of the generated meshes, we utilize a range of metrics, including Method, Average Test Set Loss, Minimum Included Angle, Maximum Included Angle, Skewness Equiangle, Aspect Ratio, Centroid Skewness, Cell Non-Orthogonality and Mesh Validity. Additionally, the Mesh Overhead is used to assess the efficiency of mesh generation.

\begin{figure*}[htbp]
    \centering
    \captionsetup[subfloat]{font=scriptsize}
    \subfloat[TFI]{\includegraphics[width=.32\columnwidth]{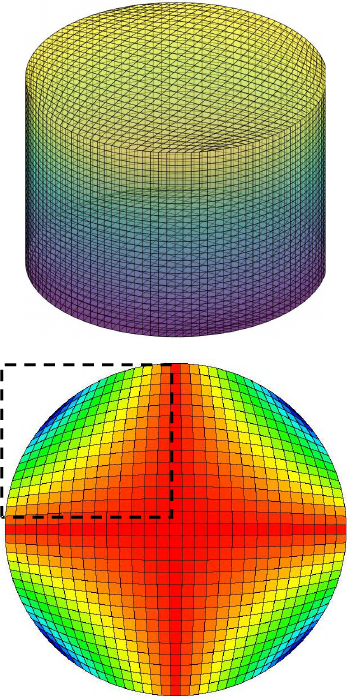}}\hspace{15pt}
    \subfloat[PDE]{\includegraphics[width=.31\columnwidth]{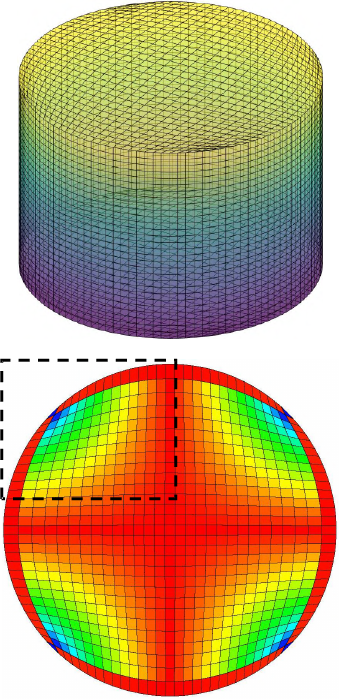}}\\
    \subfloat[PINN]{\includegraphics[width=.32\columnwidth]{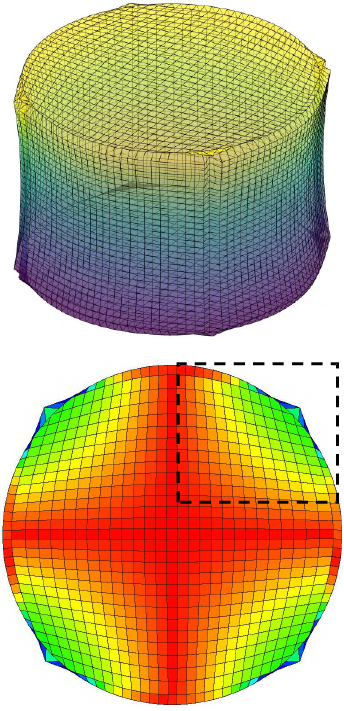}}\hspace{15pt}
    \subfloat[MGNet]{\includegraphics[width=.32\columnwidth]{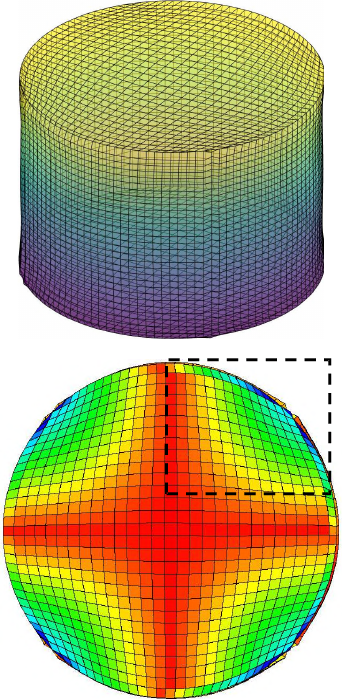}}\hspace{15pt}
    \subfloat[3DMeshNet]{\includegraphics[width=.32\columnwidth]{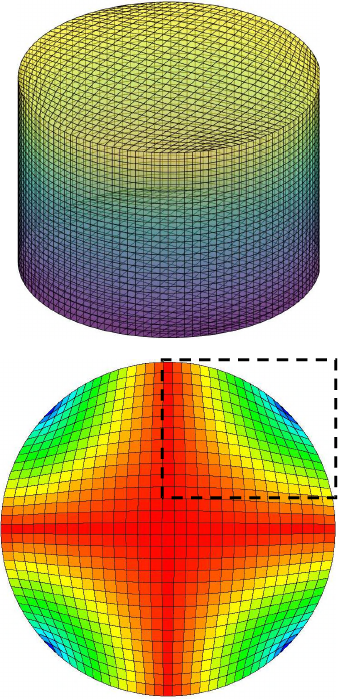}}    
    \caption{Comparison of structured mesh generation results for 3D-case1 geometry employing TFI, PDE, PINN, MGNet, and 3DMeshNet methods. The subfigures below the 3D graph are visualizations of the 3D block's cross-sections, where colors indicate mesh quality: red for higher orthogonality and blue for lower.}
    \label{3D-case1}
\end{figure*}

\begin{table*}[!t]
\centering
\caption{Numerical results comparison of structured mesh generation on 3D-case1 geometry using  TFI, PDE, PINN, MGNet and 3DMeshNet methods.}
\label{tab:3d-case1-comparison}
\begin{tabular}{lllllllll}
\hline
\thead{Method} & \thead{Avg. Test \\ Set Loss} & \thead{Min. \\ Included Angle} & \thead{Max. \\ Included Angle} & \thead{Equiangle \\Skewness } & \thead{Aspect \\ Ratio} & \thead{Centroid \\ Skewness} & \thead{Cell Non- \\ Orthogonality} & \thead{Mesh \\ Validity} \\
\hline
TFI & - & 71.29516 & 109.84139 & 0.22069 & \textbf{1.31980} & 0.11211 & 21.63536 & \multicolumn{1}{l}{Y} \\
PDE & - & 73.14630 & 108.09262 & 0.20103 & 1.37678 & 0.07211 & 18.04514 & \multicolumn{1}{l}{Y} \\
PINN & \(7.767 \times 10^{-4}\) & 69.93509 & 109.81369 & 0.22440 & 1.39929 & 0.09855 & 22.00432 & \multicolumn{1}{l}{N} \\
MGNet & \(5.652 \times 10^{-4}\) & 73.03043 & 108.03066 & 0.20105 & 1.37507& \textbf{0.07003} & 18.11415 & \multicolumn{1}{l}{Y} \\
3DMeshNet & \(\bm{1.205 \times 10^{-4}}\) & \textbf{73.19016} & \textbf{107.98604} & \textbf{0.19992} & 1.37739 & 0.07106 & \textbf{18.02890} & \multicolumn{1}{l}{Y} \\
\hline
\end{tabular}
\end{table*}

Table~\ref{tab:3d-case1-comparison} and Fig.~\ref{3D-case1} demonstrate that 3DMeshNet achieves the lowest Average Test Set Loss, surpassing PINN and MGNet. Regarding mesh quality, 3DMeshNet achieves the highest Minimum Included Angle at 73.19 and the Lowest Maximum Included Angle at 107.98, indicating superior orthogonality. It also records the lowest Skewness Equiangle value at 0.19, indicative of a shape nearing the ideal equiangular configuration. A visual analysis of Fig.~\ref{3D-case1} and its detailed views showcases the significantly inferior mesh quality at the circle's corners with the TFI method relative to 3DMeshNet. In addition, the PINN method demonstrates a limited capability to fit boundaries at corners, while 3DMeshNet exhibits excellent boundary surface adherence, enhancing the overall mesh quality.

\begin{figure*}[htbp]
    \centering
    \captionsetup[subfloat]{font=scriptsize}
    \subfloat[TFI]{\includegraphics[width=.30\columnwidth]{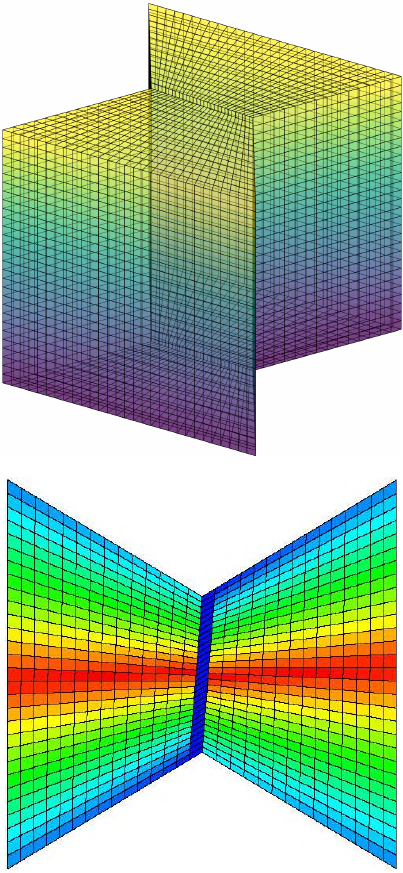}}\hspace{15pt}
    \subfloat[PDE]{\includegraphics[width=.30\columnwidth]{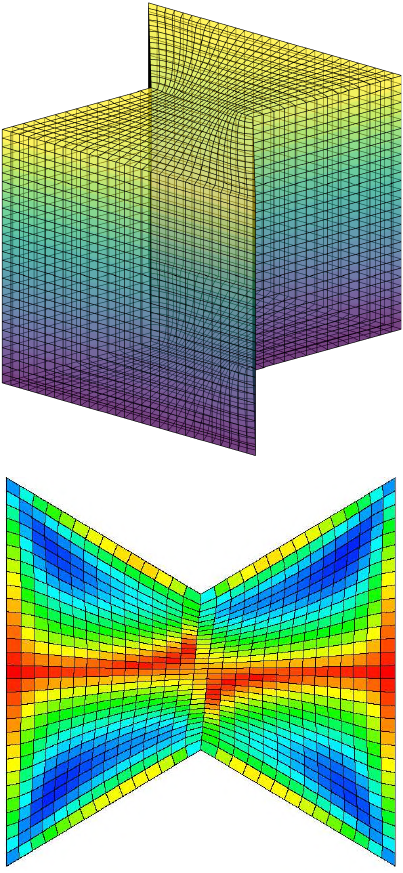}}\\
    \subfloat[PINN]{\includegraphics[width=.30\columnwidth]{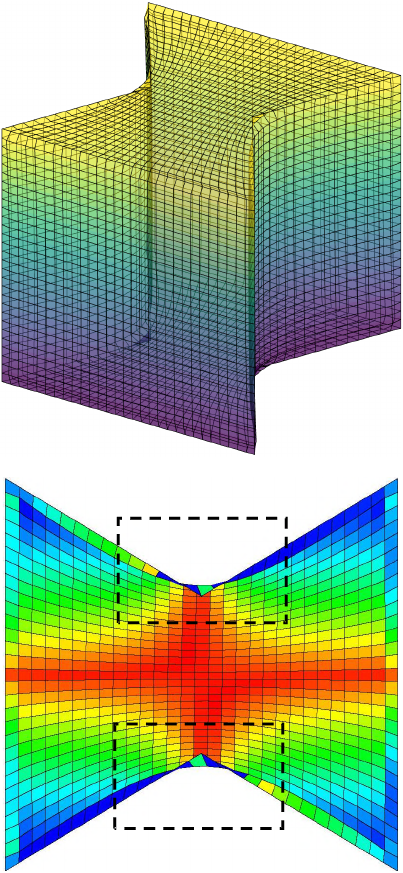}}\hspace{15pt}
    \subfloat[MGNet]{\includegraphics[width=.30\columnwidth]{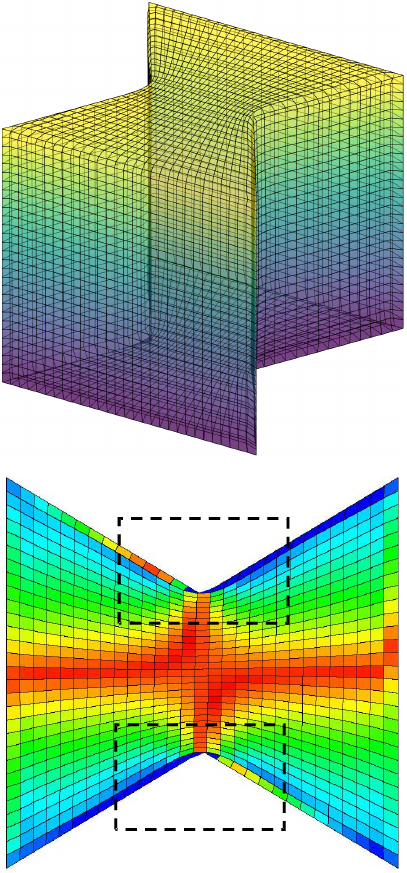}}\hspace{15pt}
    \subfloat[3DMeshNet]{\includegraphics[width=.30\columnwidth]{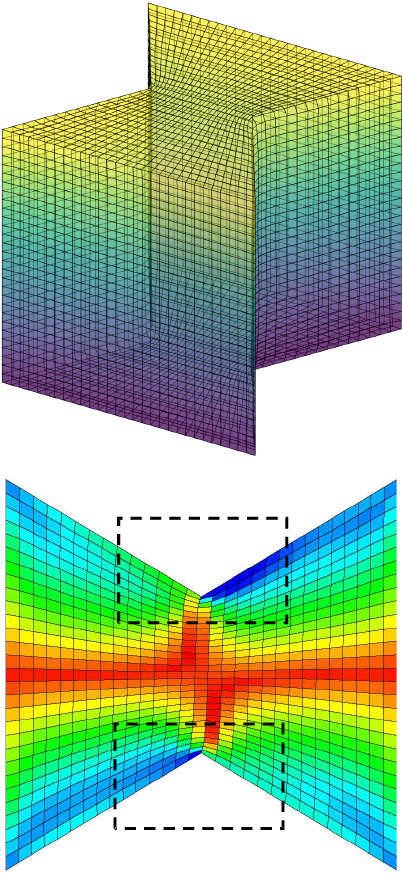}} 
    \caption{Comparison of structured mesh generation results on 3D-case2 geometry using TFI, PDE, PINN, MGNet, and 3DMeshNet methods. The subfigures below the 3D graph are visualizations of the 3D block's cross-sections, where colors indicate mesh quality: red for higher orthogonality and blue for lower.}
    \label{3D-case2}
\end{figure*}

\begin{table*}[!t]
\centering
\caption{Numerical results comparison of structured mesh generation on 3D-case2 geometry using TFI, PDE, PINN, MGNet and 3DMeshNet methods}
\label{tab:3d-case2-comparison}
\begin{tabular}{lllllllll}
\hline
\thead{Method} & \thead{Avg. Test \\ Set Loss} & \thead{Min. \\ Included Angle} & \thead{Max. \\ Included Angle} & \thead{Equiangle\\Skewness } & \thead{Aspect \\ Ratio} & \thead{Centroid \\ Skewness} & \thead{Cell Non- \\ Orthogonality} & \thead{Mesh \\ Validity} \\
\hline
TFI & - & 72.02420 & 107.98571 & 0.20151 & 1.61927 & 0.05835 & 18.30840 & Y \\
PDE & - & 73.91890 & 106.08204 & 0.17957 & 1.59924 & 0.05029 & 16.24168 & Y \\
PINN & \(9.32628 \times 10^{-5}\) & 74.10544 & 105.89219 & 0.17752 & 1.60238 & 0.04914 & 16.03306 & Y \\
MGNet & \(6.45489 \times 10^{-5}\) &74.34177  & 105.66810 &0.17471 & 1.59769& 0.04785&15.81563 & Y \\
3DMeshNet & \(\bm{1.33345 \times 10^{-5}}\) & \textbf{75.11612} & \textbf{104.90331} & \textbf{0.16605} & \textbf{1.59037} & \textbf{0.04526} & \textbf{15.06806} & Y \\
\hline
\end{tabular}
\end{table*}

Table~\ref{tab:3d-case2-comparison} and Fig.~\ref{3D-case2} show 3DMeshNet stands out with the lowest Test Set Loss at 1.33e-05,  better than PINN's 9.32e-05. In terms of angular metrics, 3DMeshNet achieves superior Minimum and Maximum Included Angles at 75.11 and 104.90, respectively, illustrating its optimal angular conformity. The Aspect Ratio of 1.59 for 3DMeshNet suggests a closer approximation to the ideal than PINN's 1.60. Furthermore, 3DMeshNet excels in metrics of Centroid Skewness and Cell Non-Orthogonality, with the lowest observed values at 0.04 and 15.06, respectively, indicating a higher fidelity of mesh quality in comparison to PINN, MGNet, TFI, and PDE.

\begin{figure*}[htbp]
    \centering
    \captionsetup[subfloat]{font=scriptsize}
    \subfloat[TFI]{\includegraphics[width=.32\columnwidth]{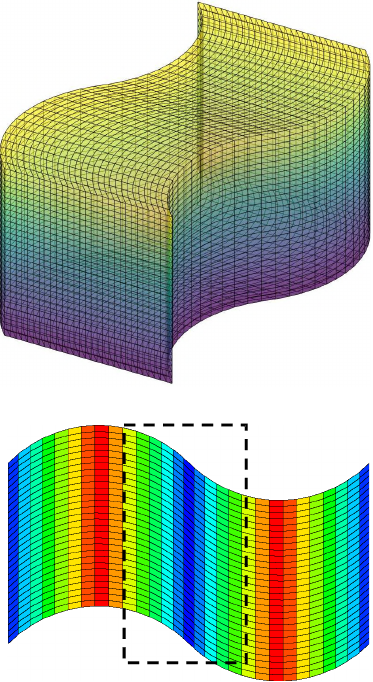}}\hspace{15pt}
    \subfloat[PDE]{\includegraphics[width=.32\columnwidth]{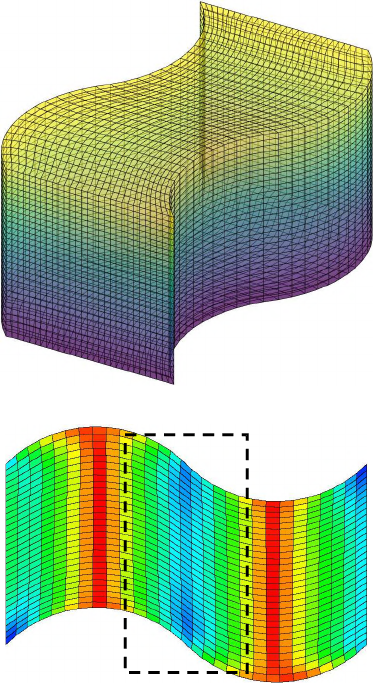}}\\
    \resizebox{0.32\columnwidth}{!}{\subfloat[PINN]{\includegraphics{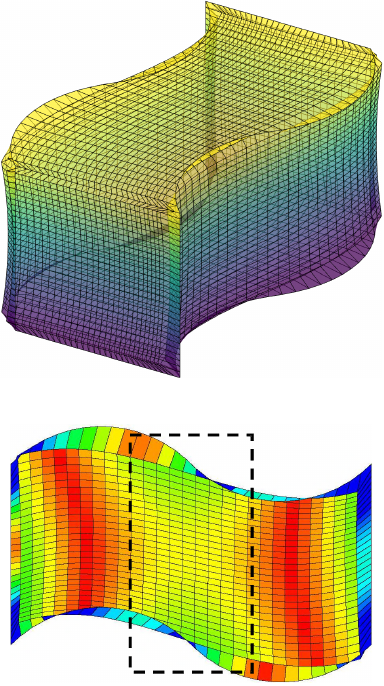}}}\hspace{15pt}
    \resizebox{0.32\columnwidth}{!}{\subfloat[MGNet]{\includegraphics{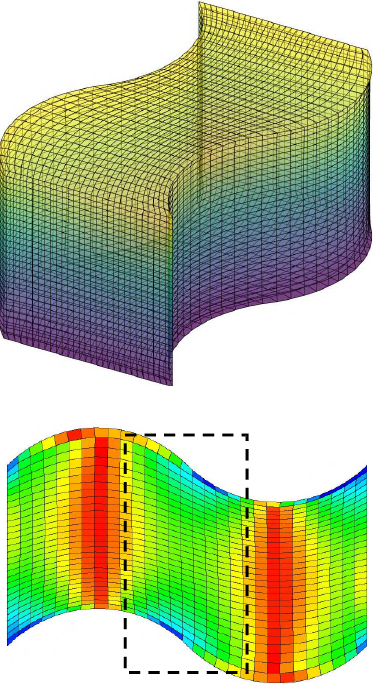}}}\hspace{15pt}
    \resizebox{0.32\columnwidth}{!}{\subfloat[3DMeshNet]{\includegraphics{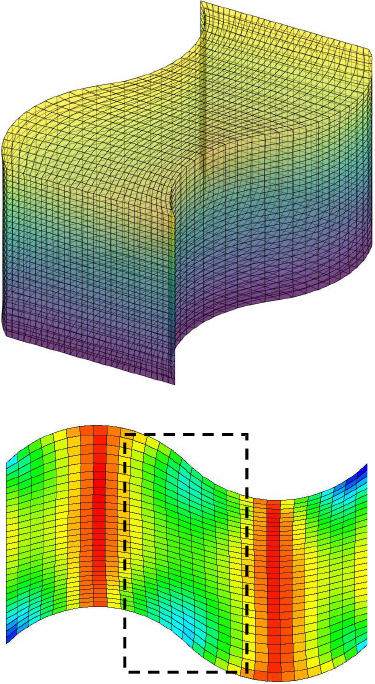}}} 
    \caption{Comparison of structured mesh generation results on 3D-case3 geometry using TFI, PDE, PINN, MGNet, and 3DMeshNet methods. The subfigures below the 3D graph are visualizations of the 3D block's cross-sections, where colors indicate mesh quality: red for higher orthogonality and blue for lower.}
    \label{3D-case3}
\end{figure*}

\begin{table*}[!t]
\centering
\caption{Numerical results comparison of structured mesh generation on 3D-case3 geometry using TFI, PDE, PINN, MGNet and 3DMeshNet methods}
\label{tab:3d-case3-comparison}
\begin{tabular}{lllllllll}
\hline
\thead{Method} & \thead{Avg. Test \\ Set Loss} & \thead{Min. \\ Included Angle} & \thead{Max. \\ Included Angle} & \thead{Equiangle\\Skewness } & \thead{Aspect \\ Ratio} & \thead{Centroid \\ Skewness} & \thead{Cell Non- \\ Orthogonality} & \thead{Mesh \\  Validity} \\
\hline
TFI & - & 67.52675 & 112.47325 & 0.24970 & \textbf{2.22038} & 0.09925 & 25.20217 &Y \\
PDE & - & 73.32343 & 107.01437 & 0.19070 & 2.31726 & 0.05531 & 18.35843 & Y\\
PINN & \(2.76692 \times 10^{-4}\) & 61.91867 & 117.90227 & 0.31438 & 3.10559 & 0.16059 & 31.22226 & N \\
MGNet & \(9.34615 \times 10^{-5}\) & 74.66378 & 105.52406 & 0.17388 & 2.30234 & 0.04624 & 16.49563 & Y \\
3DMeshNet &\(\bm{1.65482 \times 10^{-5}}\)  & \textbf{74.92872}&\textbf{105.14837} &\textbf{0.16992} & 2.29716 & \textbf{0.04418} & \textbf{16.15760} &  Y \\
\hline
\end{tabular}
\end{table*}

Table~\ref{tab:3d-case3-comparison} and Fig.~\ref{3D-case3} demonstrate that 3DMeshNet outperforms all others in Centroid Skewness and Cell Non-Orthogonality, achieving the lowest values of 0.044 and 16.15, respectively, indicating superior mesh quality. Moreover, Fig.~\ref{3D-case3} shows that 3DMeshNet enhances mesh orthogonality in the highlighted region compared to PDE and TFI methods, while the PINN method exhibits significant mesh cell distortion. MGNet performs better than PINN but is still slightly inferior to 3DMeshNet.

\begin{figure*}[htbp]
    \centering
    \captionsetup[subfloat]{font=scriptsize}
    \subfloat[TFI]{\includegraphics[width=.31\columnwidth]{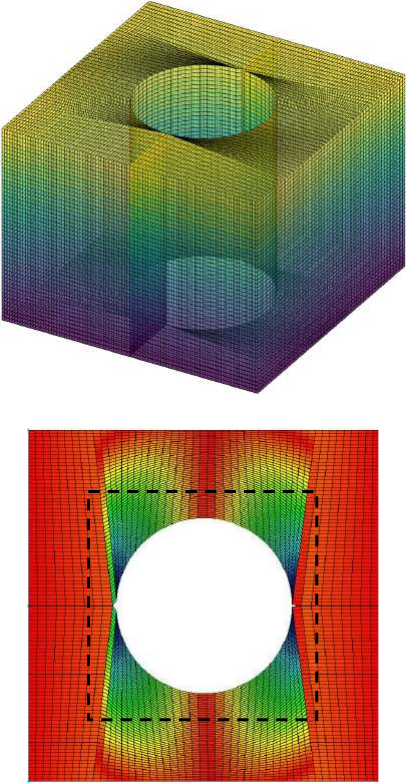}}\hspace{15pt}
    \subfloat[PDE]{\includegraphics[width=.31\columnwidth]{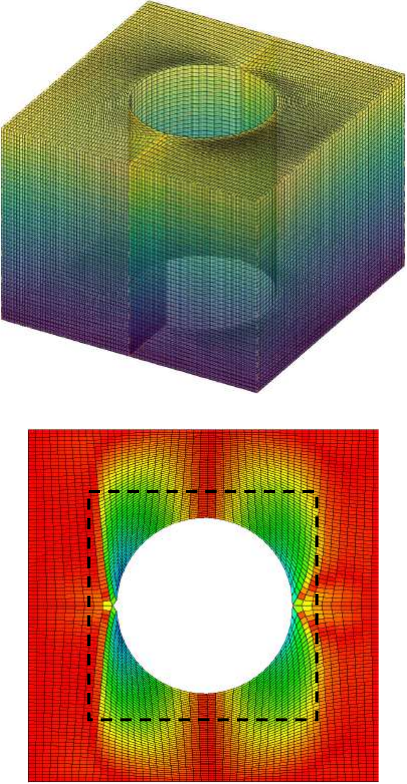}}\\
    \subfloat[PINN]{\includegraphics[width=.31\columnwidth]{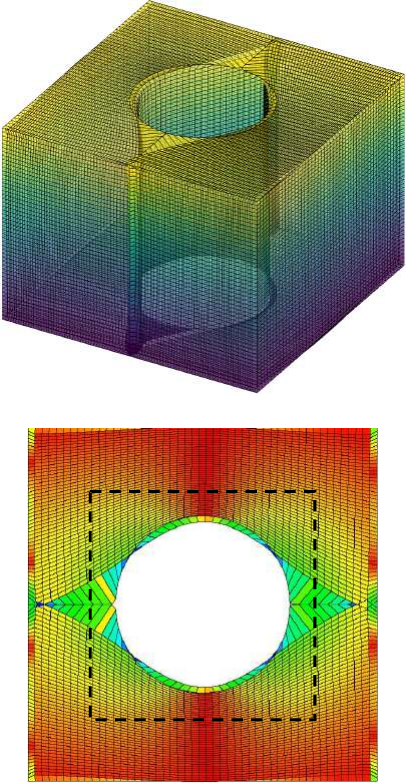}}\hspace{15pt}
    \subfloat[MGNet]{\includegraphics[width=.31\columnwidth]{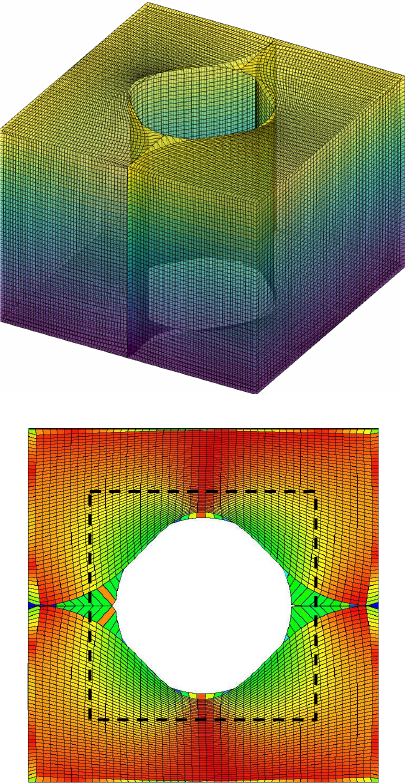}}\hspace{15pt}
    \subfloat[3DMeshNet]{\includegraphics[width=.31\columnwidth]{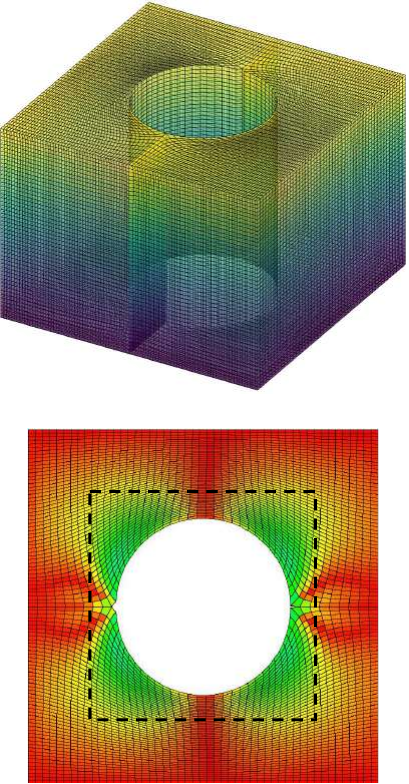}} 
    \caption{Comparison of structured mesh generation results on 3D-case4 geometry using TFI, PDE, PINN, MGNet, and 3DMeshNet methods. The subfigures below the 3D graph are visualizations of the 3D block's cross-sections, where colors indicate mesh quality: red for higher orthogonality and blue(green) for lower.}
    \label{3D-case4}
\end{figure*}

\begin{table*}[!t]
\centering
\caption{Numerical results comparison of structured mesh generation on 3D-case4 geometry using TFI, PDE, PINN, MGNet and 3DMeshNet methods}
\label{tab:3d-case4-comparison}
\begin{tabular}{lllllllll}
\hline
\thead{Method} & \thead{Avg. Test \\ Set Loss} & \thead{Min. \\ Included Angle} & \thead{Max. \\ Included Angle} & \thead{Equiangle\\Skewness } & \thead{Aspect \\ Ratio} & \thead{Centroid \\ Skewness} & \thead{Cell Non- \\ Orthogonality} & \thead{Mesh \\  Validity} \\
\hline
TFI & - & 72.60045 & 107.57008 & 0.19559 & 2.92629 & 0.09432 & 17.96599 & Y \\
PDE & - & 75.68917 & 104.43286 & 0.16128 & 2.87590 & 0.06029 & 14.77475 & Y \\
PINN & \(6.38450 \times 10^{-4}\) & \textbf{78.82539} & \textbf{101.18939} & \textbf{0.12520} & \textbf{2.75698} & \textbf{0.02872} & \textbf{11.47250} & N \\
MGNet& \(1.64892 \times 10^{-4}\)  & 76.76285 & 103.36702 & 0.14917 & 2.86600 & 0.04316 & 13.51371 & N \\
3DMeshNet& \(\bm{2.26415 \times 10^{-5}}\) &77.24883 & 102.81078 & 0.14345 &  2.82172 &  0.03892 & 13.12343 & Y \\
\hline
\end{tabular}
\end{table*}

Table~\ref{tab:3d-case4-comparison} and Fig.~\ref{3D-case4} present the Average Test Set Loss with 3DMeshNet, achieving a value of 2.26e-05, representing an order of magnitude improvement over PINN's 6.38e-4. Furthermore, 3DMeshNet outperforms TFI, PDE and MGNet by achieving minimum and maximum included angles of 77.24 and 102.81, respectively. In contrast, the mesh generated by PINN is invalid. 3DMeshNet accurately captures the object's surface shape while maintaining favorable included angles, resulting in higher quality mesh generation than PINN, as illustrated in Fig.~\ref{3D-case4}.

\subsection{Meshing Overhead}
Here, we compare 3DMeshNet's meshing overhead with TFI, PDE, PINN, MGNet and 3DMeshNet methods to validate 3DMeshNet's applicability in engineering applications as a reliable mesh generator. We test different mesh sizes ranging from \(30^2\) to \(100^3\) to evaluate the impact of mesh density on the mesh generation systems. In Table~\ref{tab:meshing-results}, we present the performance of the various methods based on execution times. All meshing processes are single-threaded in Python on the Intel Core i5-1240P  CPU, and each meshing operation is repeated ten times. The average time is recorded as the final overhead.

\begin{table*}[htbp]
\centering
\caption{Meshing Overhead results in 2D and 3D datasets. The mesh size of 2D-case1 is \(30^2\), 2D-case2 is \(282^2\); 3D-case1, 3D-case2, 3D-case3 is \(30^3\), 3D-4 is \(100^3\).}
\label{tab:meshing-results}
\begin{tabular}{lllllll}
\hline
Method      & 2D-case1      & 2D-case2       & 3D-case1      & 3D-case2       & 3D-case3       & 3D-case4\\ \hline
TFI         & 0.58299   & 1.17128 & 0.43193
   & 0.22365 & 0.52237    & 14.67969 \\
PDE         & 187.83363 & 1623.18041  & 828.05571 & 739.08905  & 767.27168  & 27328.36674 \\
PINN        & 0.00449 & 0.00349 & 0.07862 & \textbf{0.05518}    & 0.06479    &   2.38957  \\ 
MGNet & \textbf{0.00399}   &  0.00449   & \textbf{0.07583} & 0.05984 & \textbf{0.06339} & 2.41105 \\ 
3DMeshNet   & 0.00499   & \textbf{0.00346}    & 0.08072 & 0.05884 & 0.06682 & \textbf{2.27391}\\
\hline
\end{tabular}
\end{table*}

\begin{table*}[htbp]
\centering
\caption{Meshing Overhead results in 2D and 3D datasets. The mesh size of 2D-case1 is \(30^2\), 2D-case2 is \(282^2\); 3D-case1, 3D-case2, 3D-case3 is \(30^3\), 3D-4 is \(100^3\).}
\label{tab:meshing-results1}
\begin{tabular}{lllll}
\hline
Method      & 3D-case1      & 3D-case2       & 3D-case3       & 3D-case4\\ \hline
TFI         &0.4319353
   & 0.22365 & 0.52236    & 14.67969 \\
PDE         & 828.05571 & 739.08905  & 767.27168  & 27328.36674 \\
PINN        & 0.07862 & 0.05518    & 0.06479    &   2.38957  \\ 
MGNet & 0.07583 & 0.05984 & 0.06339 & 2.41105 \\ 
3DMeshNet   & 0.08072 & 0.05884 & 0.06682 & 2.27391\\
\hline
\end{tabular}
\end{table*}

\begin{table*}[htbp]
\centering
\caption{Meshing Overhead results in 2D and 3D datasets. The mesh size of 2D-case1 is \(30^2\), 2D-case2 is \(282^2\); 3D-case1, 3D-case2, 3D-case3 is \(30^3\), 3D-4 is \(100^3\).}
\label{tab:meshing-results1}
\begin{tabular}{lllll}
\hline
Method      & 3D-case1      & 3D-case2       & 3D-case3       & 3D-case4\\ \hline
TFI         & 0.43193

   & 0.22365 & 0.52236    & 14.67969 \\
PDE         & 828.05571 & 739.08905  & 767.27168  & 27328.36674 \\
3DMeshNet   & 0.08072 & 0.05884 & 0.06682 & 2.27391\\
\hline
\end{tabular}
\end{table*}

Generally, meshing overhead increases as mesh sizes grow larger. TFI exhibits linear overhead scaling with mesh size. In 2D cases, a 7.9x increase in mesh scale leads to an approximately 8.5x rise in generation time. In three-dimensional scenarios, a 37x increase in mesh scale corresponds to a 34x increase in generation time. However, the algebraic interpolation method frequently yields skewed or distorted meshes along complex boundary curves, potentially hindering convergence or causing numerical issues during numerical analysis.

 Regarding the PDE method, overhead computation involves a finite number of elliptic smoothing iterations. Results indicate that the PDE method needs more computational time, primarily due to the expensive numerical iterations of the elliptic PDE system. It takes less than 27 minutes for a mesh size of \(300^2\) and approximately 7 hours for the largest size of \(100^3\). The PDE method's execution time significantly surpasses that of algebraic and neural network-based methods.

Table~\ref{tab:meshing-results} suggests that 3DMeshNet effectively balances mesh quality and meshing overhead, generating high-quality meshes at a lower cost than the PDE method. Its near-linear scaling of overhead and independent feed-forward prediction make parallelization easy on modern systems, leading to faster execution times. Furthermore, compared to PINN and MGNet, 3DMeshNet reduces training time by 85\%. Consequently, 3DMeshNet demonstrates outstanding performance in both training time and meshing overhead.

\subsection{Ablation experiments}

This section presents ablation experiments evaluating mesh generation on 2D and 3D datasets and training time assessment. Our experiments utilize models built upon the PINN architecture as the backbone. FD-PINN substitutes the AD layer in PINN with an FD layer (detailed in Section \ref{FDMethod}). ReWeight-GP-PINN combines the loss function reweighting and gradient projection mechanisms described in Sections \ref{LossFunctionReweighting} and \ref{Gradient Projection}, respectively, into the PINN backbone. Lastly, 3DMeshNet is the model proposed in this paper.

\begin{table*}[!t]
\centering
\caption{The ablation experiments on various 2D and 3D datasets evaluate the metrics of maximum and minimum angles. The optimal mesh quality is highlighted in bold.}
\label{tab:ablation-experiments}
\begin{tabular}{lllllll}
\toprule
\thead{Dataset} & \multicolumn{2}{c}{\thead{2D}} & \multicolumn{4}{c}{\thead{3D}} \\
\cmidrule(lr){2-3} \cmidrule(lr){4-7}
& \makecell{Case3} & \makecell{Case4} & \makecell{Semi-\\cylindrical} & \makecell{Topcurve} & \makecell{Semicircular\\ canal} & \makecell{Hua} \\
\midrule
PINN            & \makecell{\textbf{77.33497}/\\103.02897} & \makecell{78.21148/\\101.28769} & \makecell{74.07149/\\107.08219} & \makecell{77.78271/\\102.20201} & \makecell{74.27333/\\106.81096} & \makecell{\textbf{74.36018}\textbf{/}\\\textbf{106.49508}} \\
FD-PINN         & \makecell{77.22772/\\103.12290} & \makecell{52.96444/\\127.06824} & \makecell{69.70034/\\111.33765} & \makecell{77.49883/\\102.37844} & \makecell{71.36079/\\109.39217} & \makecell{58.80061/\\122.19004} \\
ReWeight-GP-PINN & \makecell{77.33131/\\ \textbf{103.01870}} & \makecell{79.63152/\\100.54142} & \makecell{73.93163/\\107.09749} & \makecell{\textbf{79.98722/}\\ \textbf{100.05296}} & \makecell{74.10959/\\106.93629} & \makecell{68.65815/\\112.98135} \\
3DMeshNet   & \makecell{77.32803/\\ 103.02407} & \makecell{\textbf{79.75287/}\\ \textbf{100.28759}} & \makecell{\textbf{75.52873/}\\ \textbf{105.73708}} & \makecell{78.76921/\\101.24903} & \makecell{\textbf{75.67834/}\\ \textbf{105.58628}} & \makecell{68.61361/\\ 112.97824} \\
\bottomrule
\end{tabular}
\end{table*}

\begin{figure*}[htbp]
    \centering
    \captionsetup[subfloat]{font=scriptsize}
    \subfloat[PINN]{\includegraphics[width=.32\columnwidth]{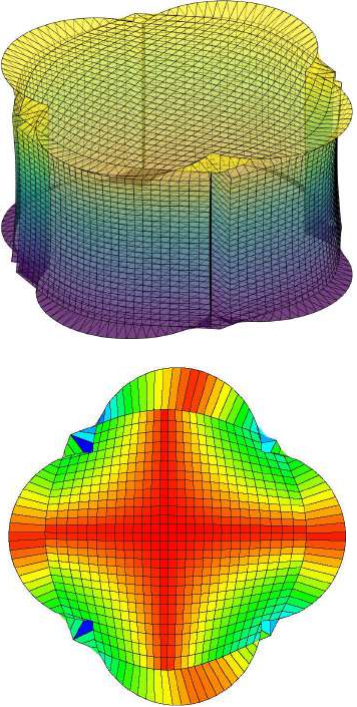}}\hspace{7pt}
    \subfloat[FD-PINN]{\includegraphics[width=.32\columnwidth]{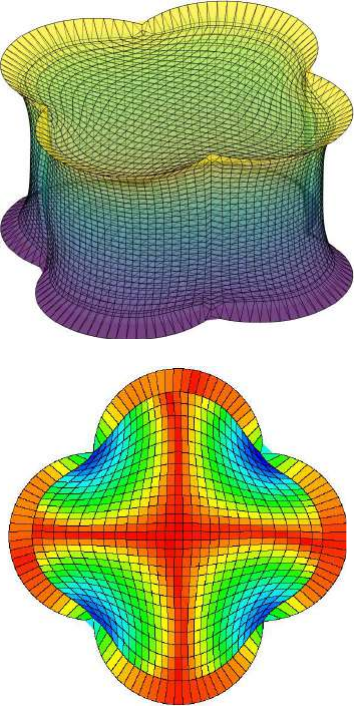}}\hspace{7pt}
    \subfloat[ReWeight-GP-PINN]{\includegraphics[width=.32\columnwidth]{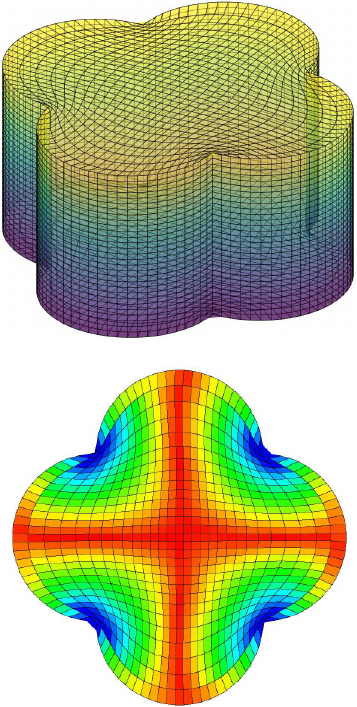}}\hspace{7pt}
    \subfloat[3DMeshNet]{\includegraphics[width=.32\columnwidth]{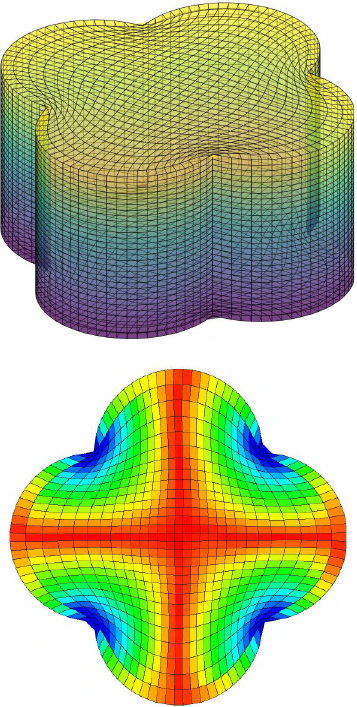}}\hspace{7pt}
    \caption{Numerical results of ablation experiments on the 3D-Hua dataset show that among the four models tested, the mesh validity for meshes generated by PINN and FD-PINN is 'N,' while ReWeight-GP-PINN and 3DMeshNet is 'Y'. In the ReWeight-GP-PINN model, both the ReWeight and Gradient Projection mechanisms are employed to enhance the model's loss function, thus combining both mechanisms for experimentation.}
    \label{3D-xiaorong}
\end{figure*}

As depicted in Table~\ref{tab:ablation-experiments}, in various instances across both 2D and 3D datasets, PINNs demonstrate superior performance on the Hua dataset. However, the Mesh Validity of PINN is 'N.' Moreover, PINN's performance on other datasets is generally mediocre, indicating that PINN, as a backbone, struggles to handle mesh generation tasks.

FD-PINN reduces computational overhead, as empirical evidence suggests it can decrease training time by 85\%. However, it fails to learn shape features, resulting in poor performance in many cases regarding Min. Included Angle and Max. Included Angle, as observed in Fig.~\ref{3D-xiaorong} b-plot, where FD-PINN generated meshes exhibit inward concavity. Nevertheless, introducing FD methods reduces computational overhead, accelerating training and model convergence.

\begin{figure}[!t]
\centering
\includegraphics[width=3.5in]{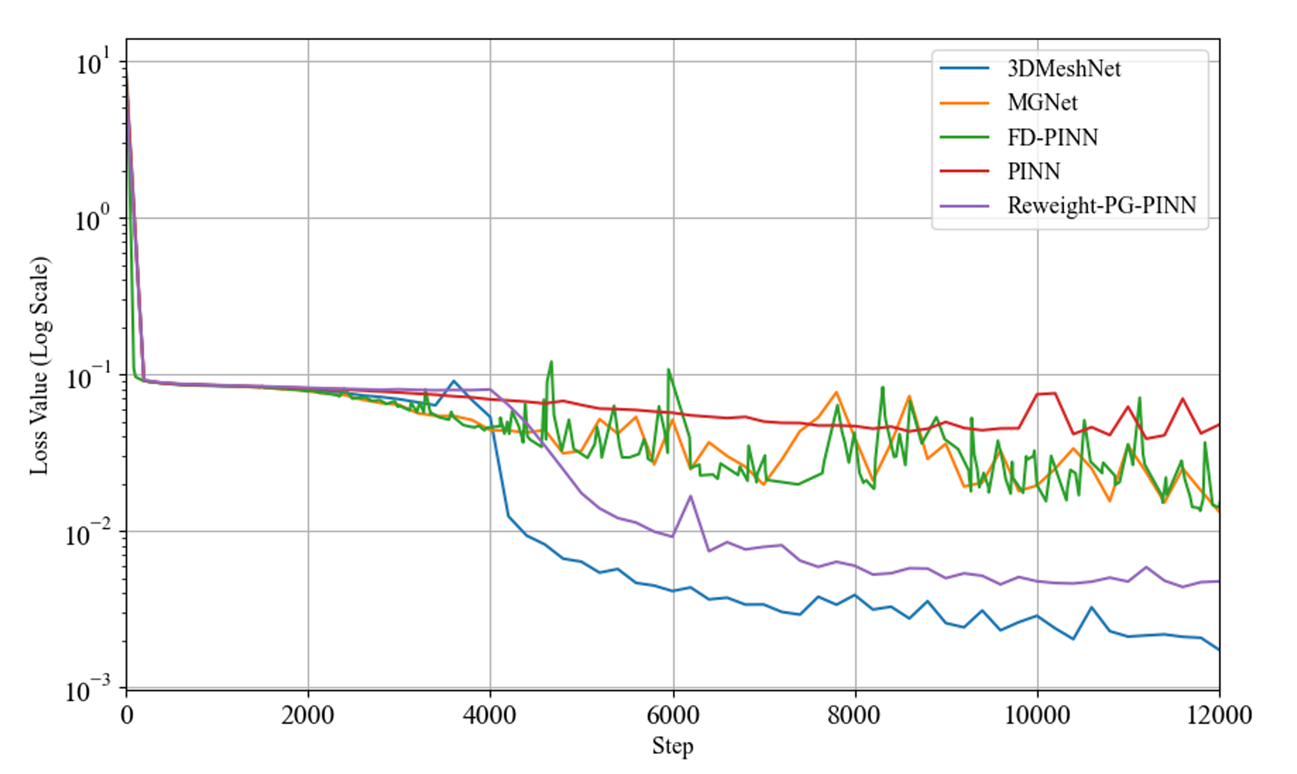}
\caption{Comparison of Average Test Set Loss on PINN, MGNet, FD-PINN, Reweight-PG-PINN and 3DMeshNet.}
\label{loss}
\end{figure}

ReWeight-GP-PINN and 3DMeshNet generate meshes with good orthogonality, as evidenced by numerical results. Additionally, when combined with Figure~\ref{loss}, it demonstrates that Loss Function Reweight and Gradient Projection can better balance PDE Loss and surface fitting Loss. However, 3DMeshNet outperforms the models above, exhibiting superior performance and faster training speed, resulting in an 85\% reduction in training time. 3DMeshNet achieves a trade-off between training time and mesh quality.

Figure~\ref{loss} depicts the loss function values for PINN, MGNet, FD-PINN, Reweight-PG-PINN, and 3DMeshNet across multiple training steps. The PINN model exhibits initially high loss values and inadequate convergence. FD-PINN and MGNet show comparable convergence, but their loss curves exhibit significant fluctuations. ReWeight-PG-PINN's loss is only slightly inferior to that of 3DMeshNet, with both curves exhibiting relatively minor fluctuations and stable convergence.

\section{Conclusion}
\label{Conclusion}
While deep learning has been applied to various aspects of numerical analysis, its utilization in structured mesh generation is an emerging field with considerable potential. We introduce 3DMeshNet, a neural network-based differential technique for creating 3D structured meshes. 3DMeshNet takes surface points on geometric shapes as input to learn the potential mapping between parametric and computational domains via a neural network and outputs a 3D structured mesh with a user-defined number of cells to partition the domain. The network approximates the mapping solution by minimizing the weighted residuals of the governing equation and surface constraint terms. Mesh point calculation is executed through feedforward prediction using deep neural networks, thereby eliminating the substantial overhead of solving PDEs. By incorporating loss function reweighting, gradient projection, and FD methods, we fasten the convergence of the network and accelerate training. Experiments show that 3DMeshNet is quick and robust, surpassing current advanced neural network-based generators and producing quality meshes equal to or greater than traditional methods. 3DMeshNet significantly reduces training time by up to 85\% compared to other neural network-based methods, while also achieving mesh overheads that are 4 to 8 times lower than those produced by the TFI method.

In our future work, we will focus on research to enhance the learning ability of 3DMeshNet, improve the generalization of 3DMeshNet, and generate high-quality 3D structured mesh on more complex shapes.

\section*{Acknowledgments}
This research work was supported in part by the National Key Research and Development Program of China: 2021YFB0300101 and the Youth Foundation of National University of Defense Technology ZK2023-11.

\bibliographystyle{IEEEtran}
\bibliography{myref}

\begin{thebibliography}{10}
\providecommand{\url}[1]{#1}
\csname url@samestyle\endcsname
\providecommand{\newblock}{\relax}
\providecommand{\bibinfo}[2]{#2}
\providecommand{\BIBentrySTDinterwordspacing}{\spaceskip=0pt\relax}
\providecommand{\BIBentryALTinterwordstretchfactor}{4}
\providecommand{\BIBentryALTinterwordspacing}{\spaceskip=\fontdimen2\font plus
\BIBentryALTinterwordstretchfactor\fontdimen3\font minus \fontdimen4\font\relax}
\providecommand{\BIBforeignlanguage}[2]{{%
\expandafter\ifx\csname l@#1\endcsname\relax
\typeout{** WARNING: IEEEtran.bst: No hyphenation pattern has been}%
\typeout{** loaded for the language `#1'. Using the pattern for}%
\typeout{** the default language instead.}%
\else
\language=\csname l@#1\endcsname
\fi
#2}}
\providecommand{\BIBdecl}{\relax}
\BIBdecl

\bibitem{chawner2019progress}
J.~R. Chawner and N.~J. Taylor, ``Progress in geometry modeling and mesh generation toward the cfd vision 2030,'' in \emph{AIAA Aviation 2019 Forum}, 2019, p. 2945.

\bibitem{hayase2015numerical}
T.~Hayase, ``Numerical simulation of real-world flows,'' \emph{Fluid Dynamics Research}, vol.~47, no.~5, p. 051201, 2015.

\bibitem{chen2021mve}
X.~Chen, J.~Liu, C.~Gong, S.~Li, Y.~Pang, and B.~Chen, ``Mve-net: An automatic 3-d structured mesh validity evaluation framework using deep neural networks,'' \emph{Computer-Aided Design}, vol. 141, p. 103104, 2021.

\bibitem{slotnick2014cfd}
J.~P. Slotnick, A.~Khodadoust, J.~Alonso, D.~Darmofal, W.~Gropp, E.~Lurie, and D.~J. Mavriplis, ``Cfd vision 2030 study: a path to revolutionary computational aerosciences,'' Tech. Rep., 2014.

\bibitem{owen2015evaluation}
S.~J. Owen and T.~R. Shelton, ``Evaluation of grid-based hex meshes for solid mechanics,'' \emph{Engineering with Computers}, vol.~31, pp. 529--543, 2015.

\bibitem{logg2009efficient}
A.~Logg, ``Efficient representation of computational meshes,'' \emph{International Journal of Computational Science and Engineering}, vol.~4, no.~4, pp. 283--295, 2009.

\bibitem{chen2022improved}
X.~Chen, J.~Liu, J.~Yan, Z.~Wang, and C.~Gong, ``An improved structured mesh generation method based on physics-informed neural networks,'' \emph{arXiv preprint arXiv:2210.09546}, 2022.

\bibitem{chen2022mgnet}
X.~Chen, T.~Li, Q.~Wan, X.~He, C.~Gong, Y.~Pang, and J.~Liu, ``Mgnet: a novel differential mesh generation method based on unsupervised neural networks,'' \emph{Engineering with Computers}, vol.~38, no.~5, pp. 4409--4421, 2022.

\bibitem{allen2008towards}
C.~Allen, ``Towards automatic structured multiblock mesh generation using improved transfinite interpolation,'' \emph{International Journal for Numerical Methods in Engineering}, vol.~74, no.~5, pp. 697--733, 2008.

\bibitem{turner2017high}
M.~Turner, ``High-order mesh generation for cfd solvers,'' Ph.D. dissertation, Imperial College London, 2017.

\bibitem{obiols2020cfdnet}
O.~Obiols-Sales, A.~Vishnu, N.~Malaya, and A.~Chandramowliswharan, ``Cfdnet: A deep learning-based accelerator for fluid simulations,'' in \emph{Proceedings of the 34th ACM international conference on supercomputing}, 2020, pp. 1--12.

\bibitem{lu2019deeponet}
L.~Lu, P.~Jin, and G.~E. Karniadakis, ``Deeponet: Learning nonlinear operators for identifying differential equations based on the universal approximation theorem of operators,'' \emph{arXiv preprint arXiv:1910.03193}, 2019.

\bibitem{jilani2009adaptive}
H.~Jilani, A.~Bahreininejad, and M.~Ahmadi, ``Adaptive finite element mesh triangulation using self-organizing neural networks,'' \emph{Advances in Engineering Software}, vol.~40, no.~11, pp. 1097--1103, 2009.

\bibitem{liu2023ispliter}
Z.~Liu, S.~Chen, X.~Gao, X.~Zhang, C.~Gong, C.~Xu, and J.~Liu, ``Ispliter: an intelligent and automatic surface mesh generator using neural networks and splitting lines,'' \emph{Advances in Aerodynamics}, vol.~5, no.~1, pp. 1--25, 2023.

\bibitem{soman2023faster}
S.~Soman and N.~Mehendale, ``Faster and efficient tetrahedral mesh generation using generator neural networks for 2d and 3d geometries,'' \emph{Neural Computing and Applications}, pp. 1--9, 2023.

\bibitem{lintermann2021computational}
A.~Lintermann, ``Computational meshing for cfd simulations,'' \emph{Clinical and biomedical engineering in the human nose: A computational fluid dynamics approach}, pp. 85--115, 2021.

\bibitem{sarrate2014unstructured}
J.~Sarrate~Ramos, E.~Ruiz-Giron{\'e}s, and F.~J. Roca~Navarro, ``Unstructured and semi-structured hexahedral mesh generation methods,'' \emph{Computational technology reviews}, vol.~10, pp. 35--64, 2014.

\bibitem{bern2000mesh}
M.~W. Bern and P.~E. Plassmann, ``Mesh generation.'' \emph{Handbook of computational geometry}, vol.~38, 2000.

\bibitem{babuska1995modeling}
I.~Babuska, \emph{Modeling, mesh generation, and adaptive numerical methods for partial differential equations}.\hskip 1em plus 0.5em minus 0.4em\relax Springer Science \& Business Media, 1995, vol.~75.

\bibitem{gawlikowski2021survey}
J.~Gawlikowski, C.~R.~N. Tassi, M.~Ali, J.~Lee, M.~Humt, J.~Feng, A.~Kruspe, R.~Triebel, P.~Jung, R.~Roscher \emph{et~al.}, ``A survey of uncertainty in deep neural networks,'' \emph{arXiv preprint arXiv:2107.03342}, 2021.

\bibitem{cuomo2022scientific}
S.~Cuomo, V.~S. Di~Cola, F.~Giampaolo, G.~Rozza, M.~Raissi, and F.~Piccialli, ``Scientific machine learning through physics--informed neural networks: Where we are and what’s next,'' \emph{Journal of Scientific Computing}, vol.~92, no.~3, p.~88, 2022.

\bibitem{zhang2018application}
Y.~Zhang, W.~J. Sung, and D.~N. Mavris, ``Application of convolutional neural network to predict airfoil lift coefficient,'' in \emph{2018 AIAA/ASCE/AHS/ASC structures, structural dynamics, and materials conference}, 2018, p. 1903.

\bibitem{gholami2020comparison}
A.~Gholami, H.~Bonakdari, A.~H. Zaji, and A.~A. Akhtari, ``A comparison of artificial intelligence-based classification techniques in predicting flow variables in sharp curved channels,'' \emph{Engineering with Computers}, vol.~36, pp. 295--324, 2020.

\bibitem{raissi2019physics}
M.~Raissi, P.~Perdikaris, and G.~E. Karniadakis, ``Physics-informed neural networks: A deep learning framework for solving forward and inverse problems involving nonlinear partial differential equations,'' \emph{Journal of Computational physics}, vol. 378, pp. 686--707, 2019.

\bibitem{chiu2022can}
P.-H. Chiu, J.~C. Wong, C.~Ooi, M.~H. Dao, and Y.-S. Ong, ``Can-pinn: A fast physics-informed neural network based on coupled-automatic--numerical differentiation method,'' \emph{Computer Methods in Applied Mechanics and Engineering}, vol. 395, p. 114909, 2022.

\bibitem{yan2023auxiliary}
J.~Yan, X.~Chen, Z.~Wang, E.~Zhou, and J.~Liu, ``Auxiliary-tasks learning for physics-informed neural network-based partial differential equations solving,'' \emph{arXiv preprint arXiv:2307.06167}, 2023.

\bibitem{wang2022evaluating}
Z.~Wang, X.~Chen, T.~Li, C.~Gong, Y.~Pang, and J.~Liu, ``Evaluating mesh quality with graph neural networks,'' \emph{Engineering with Computers}, vol.~38, no.~5, pp. 4663--4673, 2022.

\bibitem{zhang2023mqenet}
H.~Zhang, H.~Li, N.~Li, and X.~Wang, ``Mqenet: A mesh quality evaluation neural network based on dynamic graph attention,'' \emph{arXiv preprint arXiv:2309.01067}, 2023.

\bibitem{yilmaz2009particle}
A.~E. Yilmaz and M.~Kuzuoglu, ``A particle swarm optimization approach for hexahedral mesh smoothing,'' \emph{International journal for numerical methods in fluids}, vol.~60, no.~1, pp. 55--78, 2009.

\bibitem{gargallo2014surface}
A.~Gargallo-Peir{\'o}, X.~Roca, and J.~Sarrate, ``A surface mesh smoothing and untangling method independent of the cad parameterization,'' \emph{Computational mechanics}, vol.~53, pp. 587--609, 2014.

\bibitem{lowther1993density}
D.~Lowther and D.~Dyck, ``A density driven mesh generator guided by a neural network,'' \emph{IEEE transactions on magnetics}, vol.~29, no.~2, pp. 1927--1930, 1993.

\bibitem{zhang2020meshingnet}
Z.~Zhang, Y.~Wang, P.~K. Jimack, and H.~Wang, ``Meshingnet: A new mesh generation method based on deep learning,'' in \emph{International Conference on Computational Science}.\hskip 1em plus 0.5em minus 0.4em\relax Springer, 2020, pp. 186--198.

\bibitem{huang2021machine}
K.~Huang, M.~Kr{\"u}gener, A.~Brown, F.~Menhorn, H.-J. Bungartz, and D.~Hartmann, ``Machine learning-based optimal mesh generation in computational fluid dynamics,'' \emph{arXiv preprint arXiv:2102.12923}, 2021.

\bibitem{papagiannopoulos2021teach}
A.~Papagiannopoulos, P.~Clausen, and F.~Avellan, ``How to teach neural networks to mesh: Application on 2-d simplicial contours,'' \emph{Neural Networks}, vol. 136, pp. 152--179, 2021.

\bibitem{peng2022automatic}
L.~Peng, W.~Nianhua, X.~Chang, L.~Zhang, and W.~Yadong, ``An automatic isotropic/anisotropic hybrid grid generation technique for viscous flow simulations based on an artificial neural network,'' \emph{Chinese Journal of Aeronautics}, vol.~35, no.~4, pp. 102--117, 2022.

\bibitem{kim2023gmr}
M.~Kim, J.~Lee, and J.~Kim, ``Gmr-net: Gcn-based mesh refinement framework for elliptic pde problems,'' \emph{Engineering with Computers}, pp. 1--17, 2023.

\bibitem{chen2023developing}
X.~Chen, J.~Liu, Q.~Zhang, J.~Liu, Q.~Wang, L.~Deng, and Y.~Pang, ``Developing a novel structured mesh generation method based on deep neural networks,'' \emph{Physics of Fluids}, vol.~35, no.~9, 2023.

\bibitem{gordon1973construction}
W.~J. Gordon and C.~A. Hall, ``Construction of curvilinear co-ordinate systems and applications to mesh generation,'' \emph{International Journal for Numerical Methods in Engineering}, vol.~7, no.~4, pp. 461--477, 1973.

\bibitem{thompson1985numerical}
J.~F. Thompson, Z.~U. Warsi, and C.~W. Mastin, \emph{Numerical grid generation: foundations and applications}.\hskip 1em plus 0.5em minus 0.4em\relax Elsevier North-Holland, Inc., 1985.

\bibitem{kendall2018multi}
A.~Kendall, Y.~Gal, and R.~Cipolla, ``Multi-task learning using uncertainty to weigh losses for scene geometry and semantics,'' in \emph{Proceedings of the IEEE conference on computer vision and pattern recognition}, 2018, pp. 7482--7491.

\end{thebibliography}
\end{document}